\documentclass[%
 aip,
 amsmath,amssymb,
 reprint,%
]{revtex4-1}

\usepackage{graphicx}
\usepackage{dcolumn}
\usepackage{bm}
\usepackage[utf8]{inputenc}
\usepackage[T1]{fontenc}
\usepackage{mathptmx}
\usepackage{etoolbox}
\usepackage{float}
\usepackage{gensymb}
\usepackage{xcolor}
\makeatletter
\def\@email#1#2{%
 \endgroup
 \patchcmd{\titleblock@produce}
  {\frontmatter@RRAPformat}
  {\frontmatter@RRAPformat{\produce@RRAP{*#1\href{mailto:#2}{#2}}}\frontmatter@RRAPformat}
  {}{}
}%
\makeatother
\begin{document}

\preprint{AIP/123-QED}

\title{Elliptical liquid jets in a supersonic cross-flow: Influence of $J$ on atomization mechanism and unsteadiness}
\author{Chandrasekhar Medipati}
 
 \affiliation{Interdisciplinary Centre for Energy Research, Indian Institute of Science - Bangalore, India}
 \affiliation{%
 Propulsion and Power, Faculty of Aerospace Engineering, Delft University of Technology, The Netherlands 
}%
\author{Sivakumar Deivandren }%
\affiliation{ 
Department of Aerospace Engineering, Indian Institute of Science - Bangalore, India
}%

\author{Raghuraman N Govardhan}
 \email{rng@iisc.ac.in}
 \email{m.r.medipati@tudelft.nl}
\affiliation{%
Department of Mechanical Engineering, Indian Institute of Science - Bangalore, India 
}%

\maketitle

\textbf{Abstract}

 In our previous study [Medipati \textit{et al}., (2025) \textit{J. Fluid Mech}. \textbf{1014}, A34] \cite{medipati2025elliptic}, a detailed experimental investigation is performed on the elliptical liquid jets in a supersonic cross-flow ($M_{\infty}$ = 2.5), focusing on the effect of orifice aspect ratio ($AR$ = spanwise dimension/streamwise dimension) on the atomization mechanism for a fixed momentum flux ratio ($J$). In this paper, we present experimental studies that show the influence of $J$ on the jet breakup mechanism, shock structures, and unsteady interactions for each $AR$. A wide range of $J$ values (1.5 to 9.7) and three $AR$ cases (0.3, 1, and 3.3) are chosen for the study. We find that in the case of lower $J$, the jet exhibits large unsteadiness, with larger wavelength Rayleigh-Taylor (RT) waves on the windward surface. In contrast, as the $J$ increases, the unsteadiness decreases, smaller and more regular RT wavelength is formed due to the enhanced drag resulting from the reduced jet deflection. However, irrespective of $J$, in the case of $AR$ = 0.3 and 1, the primary atomization mechanism is due to the formation of Kelvin-Helmholtz instabilities (KHI) on the lateral surfaces. Furthermore, in the case of lower $J$, the shock waves formed upstream of the jet are highly corrugated with significant variations in time. The intense interaction of the liquid jet with the oncoming boundary layer streaks, in the case of lower $J$, is the primary source of large-scale unsteadiness. These findings highlight the significance of $J$ on the atomization mechanism in supersonic cross-flow.

\section{Introduction}
\label{Introduction}
\begin{figure*}
        \centering
        \includegraphics[width = 0.7\textwidth]{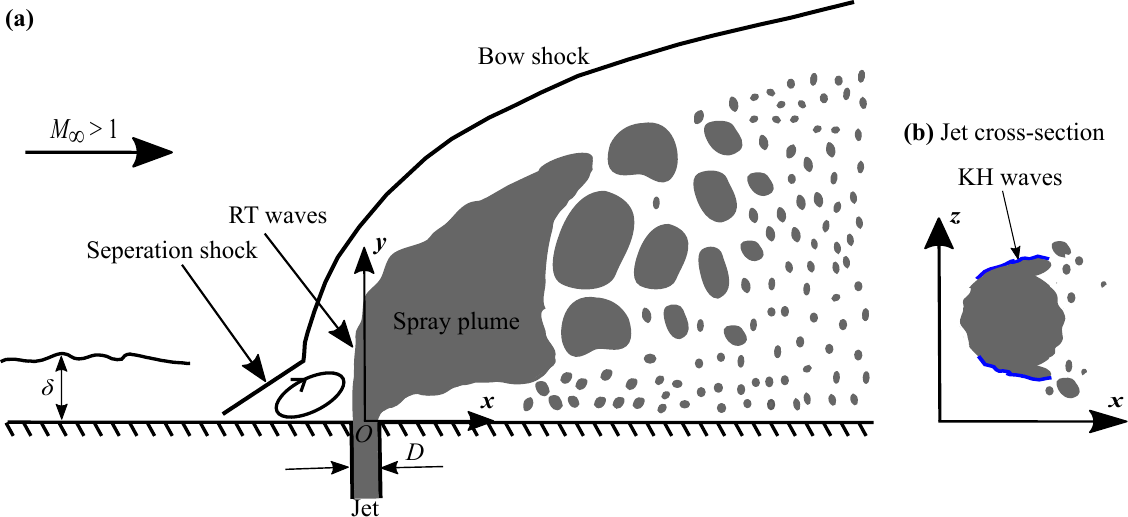}
        \caption{Schematic illustrating the main flow features of liquid jet injection into a supersonic cross-flow in the (i) transverse ($x$-$y$) and (ii) spanwise planes ($x$-$z$).}
        \label{fig:1}
\end{figure*}
\par
Transverse injection of a liquid jet into a supersonic cross-flow finds application in liquid fuel-based SCRAMJET combustors \cite{karagozian2010transverse}. Since the air entering into these combustion chambers are at supersonic speeds, the intense aerodynamic forces (pressure and shear) of the oncoming air as experienced by the injected jet results in it's rapid deflection and subsequent atomization leading to the formation of a dense spray plume with fine droplets traversing at high speeds \citep{fdida2022penetration, lin2004structures, medipati2025elliptic}. The efficiency of these engines depends on the proper mixing between the high-speed air and the fuel. The instantaneous flow field interactions and the evolution of the spray plume into the cross-flow are primarily governed by the momentum flux ratio of the injected jet and the cross-flow ($J$), cross-flow/aerodynamic Weber number ($We_\infty$), and the free stream boundary-layer thickness ($\delta$). Injection of gaseous \cite {fric1994vortical,mahesh2013interaction} and liquid jets \cite{broumand2016liquid,karagozian2010transverse} into subsonic cross-flow has been studied both experimentally and numerically by many researchers for a wide variety of configurations. Whereas the investigations pertaining to high-speed flows are limited, as reinforced by \cite{mahesh2013interaction} in his review article, where he points out that "Quantitative data at high speeds are less common, and visualizations still form an important component in estimating penetration and mixing". The present study focuses on a detailed investigation of the influence of $J$ for different orifice \textit{AR} on the mean as well as unsteady flow field characteristics, such as the instantaneous formation of surface waves on the liquid-gas interface, jet penetration height, shock structures, and the self-induced amplitude and frequencies of liquid jet pressure fluctuations. We shall see that $J$ significantly impacts the aforementioned flow field characteristics in all the orifice \textit{AR} cases. 
\par
A schematic highlighting the key flow features during the interaction of a transverse liquid jet with a supersonic flow is depicted in Fig.~\ref{fig:1}. In contrast to subsonic studies \cite{wu1997breakup,sallam2004breakup,amighi2019global,ashgriz2011handbook,prakash2018liquid,broumand2016liquid}, when a circular liquid jet is injected transversely into supersonic flow ($We_\infty=\rho_\infty U_\infty^2 D / \sigma > $ 2000), near to the stagnation region on the windward side of the jet, a rapid acceleration of the high density liquid jet by the low density high-speed air takes place. This accelerative destabilization mechanism is the well-known Rayleigh-Taylor instability (RTI) \cite{taylor1950instability}, resulting in the formation of intense RT waves\cite{medipati2025elliptic} as illustrated in Fig.\ref{fig:1}(a). Further, these RT waves grow rapidly due to the continuous acceleration of the high-speed air and eventually lead to the jet column fracture\cite{varkevisser2024effect,medipati2025elliptic}. Meanwhile, the lateral surfaces of the jet in the spanwise plane encounter Kelvin-Helmholtz instability (KHI) (Fig.\ref{fig:1}b) due to the intense shear\cite{xiao2016large,chai2015numerical} that arises because of the difference in the streamwise velocities of the low horizontal momentum liquid jet and the high-speed air \cite{behzad2016surface,xiao2016large}. This plays an important role in the amount of mass stripping from the lateral surface, which can alter the jet deflection behavior, formation of the spray plume, as well as the distribution of the droplets across the plume \cite{mashayek2008improved,broumand2016two,medipati2025elliptic}. Hence, the atomization mechanism of the transverse liquid jet in supersonic flow is an involved combination of RTI and KHI instabilities. This is analogous to the atomization of a droplet in high-speed cross-flow \cite{joseph1999breakup, engel1958fragmentation, hsiang1992near}. The complexity of this flow field is further amplified by the formation of a three-dimensional bow shock ahead of the jet due to the obstruction caused by the liquid jet to the oncoming supersonic flow. As the free-stream wall boundary layer experiences an adverse pressure gradient created by the shock wave, it leads to the separation of the boundary layer, resulting in significant shock wave boundary layer interactions (SWBLIs), which eventually lead to large-scale unsteadiness in the jet penetration height, shock curvature and position, and distribution of the droplets across the plume \cite{medipati2023liquid}. To understand the intricate fluid dynamics of SWBLI and the source of low-frequency shock unsteadiness, numerous studies have been performed in the past using various canonical solid body configurations \cite{clemens2014low,dolling200050, dussauge2008shock,ganapathisubramani2007effects, beresh2002relationship,murugan2016shock}. In this connection, Murugan and Govardhan\cite{munuswamy2022shock} have performed a detailed experimental investigation of the unsteady behavior of a sonic gaseous jet injected into a supersonic flow using simultaneous particle image velocimetry (PIV) on the jet and free-stream sides. They found that an instant corresponding to a thinner upstream boundary layer associated with a high momentum streak leads to the downstream displacement of the separation shock along with a lower jet penetration height and the downstream displacement of the bow shock wave, while the opposite effects were seen for an instant with a thicker boundary layer or low-speed streak. As anticipated, one would expect qualitatively similar dynamics in the case of a liquid jet like in the present study.     
\par
Quantifying the jet penetration height is of prime importance in any jet in cross-flow study \cite{mahesh2013interaction}. Many studies are available for a circular liquid jet in supersonic cross-flow focusing on the mean penetration height primarily as a function of $J$ for a variety of cross-flow Mach numbers and jet conditions \cite{kush1973liquid,lin2002penetration,lin2004structures,ghenai2009penetration,sathiyamoorthy2020penetration,medipati2023liquid,fdida2022penetration,medipati2025elliptic}. Further, due to the progressive advancement in optical diagnostics, the droplet size and their velocity distributions across the spray plume have been determined \cite{nejad1983effects,lin2004structures,wu2015investigations,medipati2023liquid,johnson2024digital}. It was found that the droplets formed are larger in the spray core with lower streamwise velocities because of their higher inertia. This typically results in `S' and `C' shaped trends for the Sauter mean diameter ($SMD$) and mean streamwise droplet velocity ($U$), respectively, as a function of the transverse direction of the spray plume \cite{lin2004structures,liu2016numerical,medipati2023liquid,johnson2024digital}.
\par
Recently, we have experimentally investigated elliptical liquid jets in a supersonic cross-flow \cite{medipati2025elliptic}, focusing on the influence of orifice aspect ratio ($AR$) on the atomization mechanism, shock structures, and the unsteady aspects of the flow field. The orifice $AR$ was varied (0.3 - 3.3) while maintaining the same $J$. It was found that in the case of $AR$ = 3.3, the liquid jet has smaller RT waves on the windward surface due to the rapid acceleration of the liquid jet, resulting from the larger drag forces experienced by the liquid jet (larger frontal area). Hence, the jet exhibits a quicker deflection into the cross-flow, a lower penetration height, and the upstream bow shock wave is free from corrugations, followed by a catastrophic breakup of the jet. In contrast, in the case of $AR$ = 0.3, the reduced acceleration of the liquid jet, leads to significantly larger RT waves, resulting in corrugated shock structures with significant variations in time. However, the atomization mechanism is primarily driven by the KHI on the large lateral surfaces, leading to the formation of liquid tongues, which further undergo atomization due to RTI, as observed in coaxial jet studies \cite{varga2003initial}. These disparities in the near-field interactions with $AR$ significantly influence the mean and unsteadiness in jet penetration behavior, scaling laws for jet penetration height, the amount of mass stripping from the jet surface, unsteady jet-shock-boundary layer interactions, and, finally, the dispersion and distribution of droplets across the spray plume. It is found that in the case of $AR$ = 3.3, where the atomization is primarily through RTI, the jet breakup time scales are lower, which leads to the formation of a steady spray core containing finer droplets that convect with higher velocities. In contrast, $AR$ = 0.3 case exhibited higher breakup time scales due to atomization being primarily through KHI, and the spray core is highly unsteady, with larger ligaments and droplets that move with lower velocities. This eventually altered the conventional `S' and `C'-shaped $SMD$ and $U$ trends of circular ($AR$ = 1) liquid jets as a function of the transverse direction of the spray plume.   

It is apparent that significant studies are available in the literature with varying $J$, but the primary focus of these studies has been to quantify the mean penetration height as a function of $J$ for a fixed $M_\infty$. A detailed quantitative analysis of how  $J$ influences the atomization mechanism, instantaneous formation of windward surface waves, shock structures, and unsteadiness has not been investigated so far, as also evident from the recent review of \cite{zhang2024review}. Therefore, the novelty of the present work lies in the detailed measurements and analysis of distinguishing features for different $J$. The primary focus is to understand how $J$ influences the instantaneous formation of near-field windward surface waves, the atomization mechanism, breakup time scales, shock structures, and the corresponding unsteady aspects due to the differences in interaction with the oncoming turbulent boundary layer streaks, as revealed by the PIV measurements upstream of the jet. The unsteady pressure measurements are also performed in the liquid injection line for varying $J$ cases to complement the visualizations and understand the effect of upstream low-frequency shock oscillations caused by the SWBLI on the self-induced liquid jet frequencies. These experiments are conducted for different orifice $AR$ cases using elliptical and circular orifices.     

The rest of the paper is organized as follows. In section \ref{Experimentaldetails}, we present the experimental details that include information about the experimental facility and experimental methods employed, and the chosen experimental conditions. This is followed by section \ref{Influence of $J$ on instabilities and shock structures} that discusses the influence of $J$ on the formation of windward surface waves on the liquid jet, the atomization mechanism, and the associated shock structures upstream of the injected liquid jet. The unsteady interactions between the jet-shock and boundary layer and their variations with $J$ are then presented in section~\ref{Unsteady behavior of penetration height and shock position}. The unsteady pressure measurements in the liquid injection line and the corresponding frequencies are presented in section~\ref{Unsteady pressure measurements in the liquid injection line}, followed by the summary and conclusions in section~\ref{Conclusions}.    

\section{Experimental details}
\label{Experimentaldetails}
\subsection{Experimental facility}
\label{Experimental facility}
\begin{figure*}
        \centering
        \includegraphics[width = 0.75\textwidth]{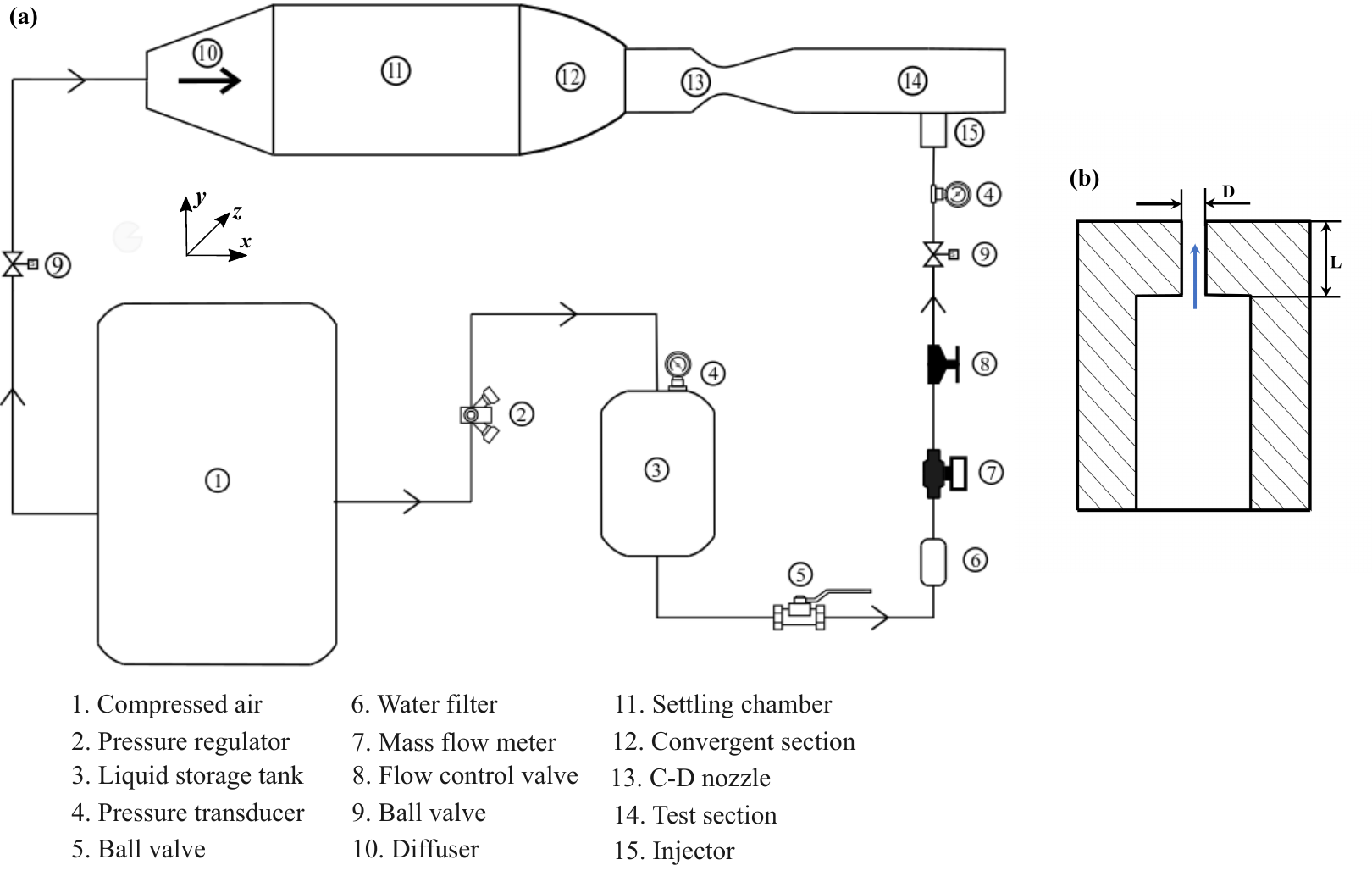}
        \caption{Schematic of the supersonic blowdown wind tunnel facility and the liquid injection system. (b) Schematic of the sharp-edged injector that is flush-mounted with respect to the tunnel bottom wall. The length and diameter of the nozzle are represented as $L$ and $D$, respectively. The direction of water flow inside the nozzle is indicated in blue.}
        \label{fig:2}
\end{figure*}
All the experimental measurements reported in this paper were conducted in an open-circuit blow-down wind tunnel at the Interdisciplinary Center for Energy Research, Indian Institute of Science, Bangalore. The experimental facility consists of a supersonic blowdown wind tunnel and a liquid injection system. The supersonic blowdown wind tunnel mainly comprises a settling chamber, a C-D nozzle, a test section, and a diffuser. In the settling chamber, a stagnation pressure of about 2.8 bar is maintained to produce a supersonic flow with a free-stream Mach number ($M_\infty$) of 2.5 in the test section for a run time of about 90 seconds. The test section has a cross-section of 15 cm x 15 cm and a length of 1 m, with both sides and the top window of the test section made transparent for optical access. The liquid nozzle is flush-mounted at 100 mm from the C-D nozzle exit, where the free-stream velocity ($U_\infty$) and the boundary layer thickness ($\delta$) were measured using PIV, and the values are found to be 585~m/s and 8.85 mm, respectively. The streamwise, cross-streamwise, and spanwise directions are represented as $x$, $y$, and $z$, respectively. The liquid injection system involves a plenum chamber that can withstand up to 70 bar, an injection line, and a sharp-edged nozzle. The source of high-pressure air for both the supersonic wind tunnel and the liquid injection system is the compressed air stored in cylindrical tanks, initially at 40 bar and 30 m³. The schematic of the wind tunnel facility and the liquid injection system is shown in Figs. \ref{fig:2} (a) and (b), respectively.

 \subsection{Experimental conditions}
\label{Experimental conditions}
\setlength{\tabcolsep}{10pt} 
\renewcommand{\arraystretch}{1.5} 
\begin{table*}
\centering
\begin{tabular}{  c   c   c   c   c   c   c   c  c}
   \hline
   Parameter &  $M_\infty$  &  $U_\infty$ (m/s)  &  $D$ (mm)  &  $U_j$ (m/s)  & $J$  &  $AR$ &  $L/D$ & $Re_j\times10^4$ \\ 
    \hline
   Values  &  2.5  &  585  &  1.2  &  16.2 - 45.5  & 1.5 - 9.7  &  0.3 - 3.3  & 2 & 2.2 - 6.2   \\
    \hline
    \end{tabular}
\caption{Values of experimental conditions chosen for the present study during jet injection.}
    \label{tab:parameters}
\end{table*}
\setlength{\tabcolsep}{10pt} 
\renewcommand{\arraystretch}{1.5} 
\begin{table*}
\centering
    \begin{tabular}{m{10mm}  >{\centering}m{25mm}  m{10mm}  m{10mm} } 
    \hline
    $AR$ & Jet exit shape & $a$ (mm) & $b$ (mm)  \\ 
   \hline
     0.3 & \includegraphics[width = 8mm,height = 3mm]{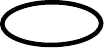} & 2.23 & 0.67  \\
     1 & \includegraphics[width = 4mm,height = 4mm]{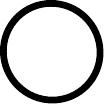} & 1.2 & 1.2  \\ 
     3.3 &\includegraphics[width = 3mm,height = 8mm]{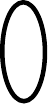} & 0.67 & 2.23 \\
      \hline
         \end{tabular}
\caption{The geometric details of the orifice used in the present study. $a$ and $b$ denote the streamwise and spanwise dimensions of the elliptical orifice, respectively.}  
     \label{tab:AR}
\end{table*}
\noindent The important dimensionless groups characterizing the flow-field are:\\
\begin{itemize}
    \item Freestream Mach number,   $M_\infty = U_\infty/\sqrt{\gamma R T_\infty}$  ,
    \item Momentum flux ratio, defined as the ratio of momentum flux of the jet to momentum flux of the cross-flow,  expressed as  $J = (\rho_j U_j^2)/(\rho_\infty U_\infty^2 )$, 
    \item Aerodynamic Weber number defined as the ratio of aerodynamic force to surface tension force,  expressed as We$_\infty=(\rho_\infty U_\infty^2 D)/\sigma$,   
    \item Reynolds number of the liquid jet  expressed as  Re$_j=(\rho_j U_j D)/\mu_j$,   
    \item Orifice aspect ratio ($AR$), defined as the ratio of spanwise dimension to streamwise dimension of the orifice ($b/a$),
\end{itemize}
\par

\noindent where $\rho_\infty$, $U_\infty$, and $\rho_j$, $U_j$  denote the density and velocity of the freestream air and the liquid jet, respectively.  The mean velocity of the liquid jet is calculated from the volume flow rate and the orifice exit area. $\sigma$ and $\mu_j$ represent surface tension of the air-water interface and the dynamic viscosity of water, respectively. The static temperature of the air, specific heat ratio, and gas constant are denoted by $T_\infty, \gamma$, and $R$, respectively. The values or range of values of experimental conditions chosen for the present investigation are summarized in Table~\ref{tab:parameters}. Since the main focus of the present study is on understanding the influence of $J$ on various characteristics of the flow field, the experiments are performed for a wide range of $J$ values (1.5 - 9.7) for each orifice aspect ratio ($AR$). Throughout the study, the cross-flow conditions are fixed corresponding to $M_\infty$ = 2.5, and the $J$ is varied purely by varying the injection velocity ($U_j$) of the water jet. The orifice $AR$ (0.3, 1, and 3.3) is changed by changing the orifice shape from circular to elliptical with two orientations with respect to cross-flow. The details of the orifice shape, orientation with respect to the cross-flow, and its geometric details are tabulated in Table~\ref{tab:AR}. It is worth noting that in the entire study, the injection location remains the same for all the experiments to ensure the liquid jet experiences a similar boundary layer thickness.    

\subsection{Experimental methods}
\label{Experimental methods}
\begin{figure*}
        \centering
        \includegraphics[width = 0.75\textwidth]{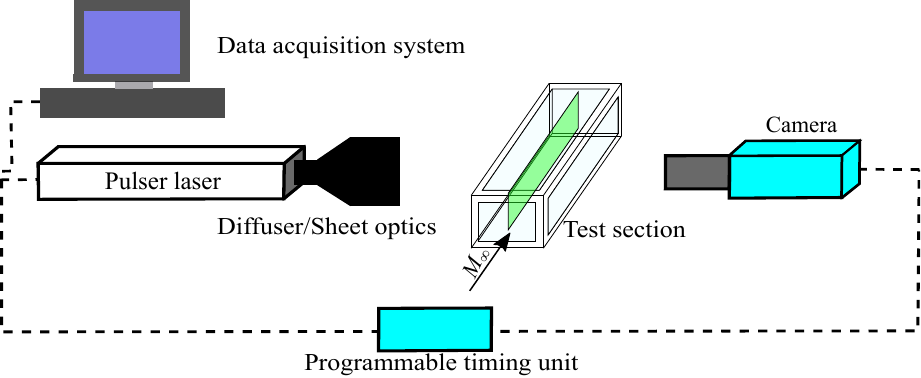}
        \caption{Schematic of the optical diagnostics used for the present study. The imaging plane is indicated in green.}
        \label{fig:3}
\end{figure*}

Atomization of a liquid jet in a supersonic flow is a transient multiphase and multiscale phenomenon\cite {fdida2022penetration}. Hence, to capture these dynamics for different $J$, we have used a variety of experimental techniques as discussed below:

Fig.\ref{fig:3} illustrates the schematic of the laser diagnostics employed for the present investigation. It comprises a nanosecond pulsed laser (Litron, 200mJ/pulse 532 nm dual cavity Nd:YAG laser), a diffuser or a sheet optics, a microsecond exposure CCD camera (Imager SX) with a resolution of 2360 x 1776 pixels, and a programmable timing unit (PTU) for synchronization. In the case of Pulsed Laser Shadowgraphy (PLS), the laser head was connected to the fluorescent diffuser, and a narrow field of view of about 30 mm x 23 mm was chosen near the jet exit. The shadowgraphy images were captured by connecting a 105 mm lens to the CCD camera. The acquisition rate of the system is limited and captured at 15 Hz. To capture the complex shock structures and the involved temporal dynamics, high-speed shadowgraphy is deployed. A key difference between the PLS and high-speed shadowgraphy is that, instead of a diffused light source, a collimated light beam is produced, and a high-speed camera (Photron, FASTCAMSA5) with a microsecond exposure setting and a field of view of about 18 mm x 24 mm is used to capture the images at 10,000 Hz. In both techniques, the light is illuminated through the back side transparent window, and the images are captured by viewing through the front side transparent window.          

Particle Image Velocimetry (PIV) was employed near the tunnel wall on the free-stream side during the injection of the liquid jet. The supersonic side is seeded with olive oil particles that act as tracer particles. These experiments were performed ahead of the liquid jet by zooming to a field of view of about 30 mm x 18 mm.  This allowed us to capture the dynamics of the oncoming turbulent boundary layer and its influence on the jet. Compared to PLS, now the laser head is replaced with a sheet optics ($f$ = -10 mm) to convert the beam into a sheet of 1.5 mm thickness. The sheet is allowed to enter the test section through the top transparent window to illuminate the entire $z$ = 0 ($x$-$y$) plane. The images are acquired with a CCD camera by viewing through the front side transparent window. These images are captured at 15 Hz by operating the entire laser system in dual pulse mode with an inter-frame time of 0.5~$\mu$s. All the measurements are performed in the mid span ($z$ = 0) as indicated with a green color in figure \ref{fig:3}.    

Unsteady pressure measurements were conducted in the liquid line to reveal the influence of $J$ on the amplitude of liquid jet fluctuations and the effect of the upstream low-frequency shock oscillations on the injected liquid jet. During these experiments, an unsteady pressure transducer (Model number: 211B5) from the Kistler group was introduced in the injection line at a location 4 meters upstream of the orifice, and measurements were performed for both with and without cross-flow conditions. These pressure traces were acquired at a sampling rate of 50 kHz using a 16-bit Data Acquisition card (model:NI 9222). The pressure time traces, spectra, and their variations with $J$ are discussed in section \ref {Unsteady pressure measurements in the liquid injection line}. Detailed information about the experimental facility, the methodology followed, and the data analysis carried out can be found in our previous works \cite{medipati2023liquid,medipati2025elliptic,munuswamy2022shock,murugan2016shock}. 
\section{Influence of $J$ on atomization mechanism and shock structures}
\label{Influence of $J$ on instabilities and shock structures}
In this section, we start by presenting the influence of $J$ on the formation of surface waves on the windward side of the jet and the initial breakup mechanism of the liquid jet. This is followed by a discussion on the jet breakup time scales and their effect on $J$. Finally, the influence of $J$ on the formation of complex shock structures upstream of the liquid jet is presented.
\subsection{Atomization mechanism}
\label{Atomization mechanism}
\begin{figure*}
        \centering
        \includegraphics[width = 0.9\textwidth]{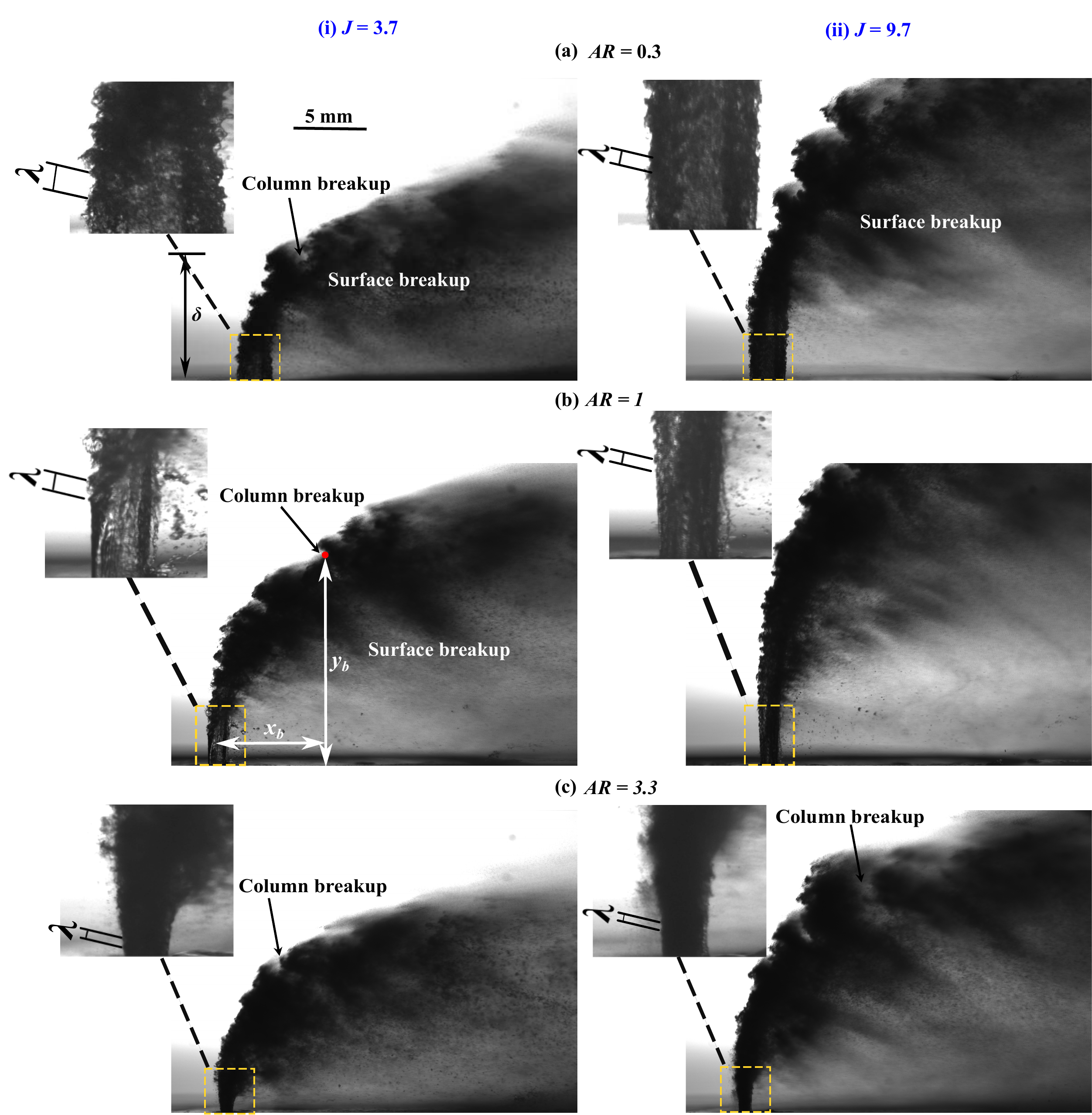}
        \caption{High-resolution instantaneous images of a water jet in a supersonic cross-flow of $M_\infty$ = 2.5, captured using pulsed laser shadowgraphy, highlighting the differences in the development of the surface waves in the windward side of the jet. These are shown for (i) $J$ = 3.7, and (ii) $J$ = 9.7, and for (a) $AR$ = 0.3, (b) $AR$ = 1 and $AR$ = 3.3, respectively. Insets on the left side of each image indicate the zoomed-in visualizations near the jet exit. The surface wavelength and the mean boundary layer thickness are represented as $\lambda$ and $\delta$, respectively. The column breakup location is indicated with a red colored dot, and its instantaneous streamwise and transverse locations from the orifice center are denoted as $x_b$ and $y_b$, respectively.}
        \label{fig:4}
\end{figure*}
Fig.\ref{fig:4} shows the microsecond exposure instantaneous visualizations of the transversely injected elliptical liquid jets in supersonic flow captured using pulsed laser shadowgraphy.  The left and right side images correspond to instantaneous visualizations for (i) $J$ = 3.7 and (ii) $J$ = 9.7, and for (a) $AR$ = 0.3, (b) $AR$ = 1, and (c) $AR$ = 3.3, respectively. A large set of these instantaneous visualizations indicate a significant influence of $J$ on the formation of the windward surface waves. As the jet penetrates into the cross-flow, the surface waves are observed to form on the windward side of the jet \citep{sallam2004breakup,varkevisser2024effect,medipati2025elliptic}. This is due to the rapid acceleration of the heavier low-speed liquid jet by the lighter high-speed air. This accelerative destabilization mechanism is the familiar Rayleigh-Taylor instability \citep{taylor1950instability}. Irrespective of orifice $AR$, in the case of lower $J$ (Fig. \ref{fig:4} (i)) and lower penetrating cases, it is evident that the windward surface waves formed are at lower transverse heights and larger in wavelengths ($\lambda$) due to the lower acceleration of the liquid jet. This is because the liquid jet experiences lower drag due to its higher deflection into the cross-flow. Furthermore, it is seen that the formation and the growth of windward surface waves and the deformation of the liquid jet between the instants are highly irregular and unsteady, followed by a sharp deflection of the jet into cross-flow with a random breakup event, resulting in the formation of an irregular liquid clump size between the instants. This is also qualitatively observed and indicated in one of the earlier works \citep{kush1973liquid}. This is most likely due to the intense interaction of the liquid jet with the oncoming boundary layer streaks, resulting from the lower penetration of the jet. This can lead to the formation of complex shock structures with significant variations in corrugations over time, as discussed in Section \ref{Shock structures}. In contrast, in the case of higher $J$ (Fig. \ref{fig:4} (ii)) and higher penetrating cases, the windward surface waves ($\lambda$) start to form at higher transverse heights and are smaller in wavelengths due to the increase in the acceleration of the liquid jet. This is because the liquid jet experiences higher drag due to lower jet deflection into the cross-flow. Moreover, the liquid jet exhibits a systematic formation and a gradual growth of windward surface waves, bending like a smooth curve, followed by a fairly regular breakup event with the formation of uniform liquid clump sizes between the instants. The differences in the drag forces acting on the liquid jet for the low and high $J$ cases are analogous to the classical problem of the flow over an inclined flat plate.  As the flat plate is aligned perpendicular to the cross-flow, the drag force experienced by the plate is highest, which is similar to the most penetrating higher $J$ case of the present study, while the drag is lowest for a flat plate aligned parallel to the cross-flow \citep{anderson2011ebook}. 

For the same $J$, as the orifice $AR$ changes, the interaction/frontal area ($A_f$) with the supersonic flow changes, and hence the drag experienced by the liquid jet also changes. In the case of $AR$ = 0.3 (Fig.~\ref{fig:4}a), irrespective of $J$, the surface waves formed are larger due to the lower acceleration experienced by the liquid jet. This is because the liquid jet experiences lower drag as a result of its lower interaction/frontal area with the oncoming supersonic air. In contrast, in the case of $AR$ = 3.3, the surface waves formed are very small due to the rapid acceleration of the liquid jet due to the larger drag experienced by it as a result of its higher frontal area.  

To summarize, the formation of RT waves on the windward liquid jet surface is a strong function of the acceleration of the liquid jet, which in turn depends on the drag experienced by the jet. In the present varying $J$ study (for a given $AR$), the acceleration of the liquid jet varies with $J$ due to the variation in the drag experienced by the liquid jet, which is caused by the variation in the jet deflection angle. In contrast, in our previous varying orifice $AR$ study (for a given $J$)\citep{medipati2025elliptic}, the acceleration of the liquid jet is also changed, but the variation is due to the variation in drag caused by the variation in the frontal area of the liquid jet (orifice $AR$). 
\begin{figure*}
        \centering
        \includegraphics[width = 0.8\textwidth]{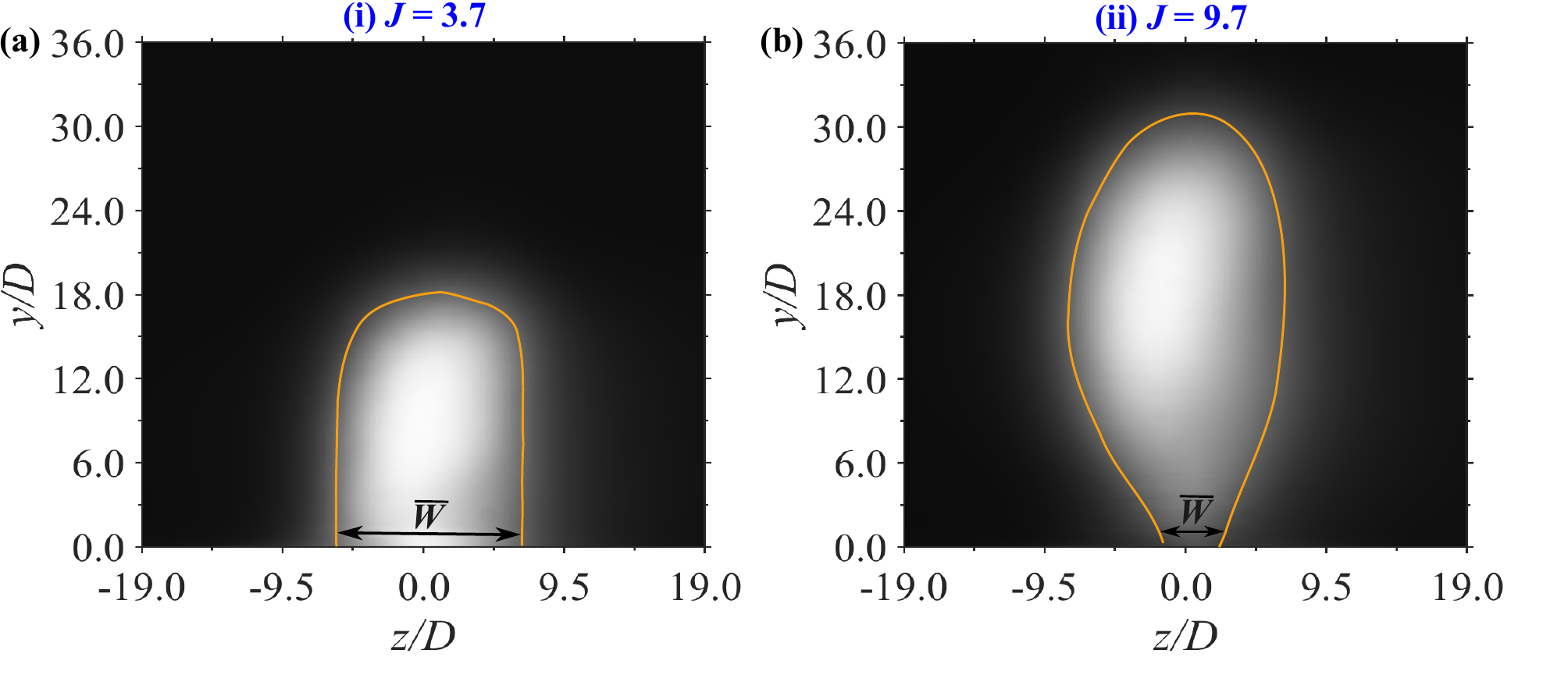}
        \caption{Influence of $J$ on the time-averaged planar laser Mie scattering images of the spray plume cross-section when a liquid jet is injected into a supersonic cross-flow. These are captured at  $x/D$ = 60 and for $AR$ = 1. The outer edge of the spray plume cross-section is shown in orange. $\overline{W}$ indicates the mean spray width in the spanwise direction.}
        \label{fig:5}
\end{figure*}

Although RT instability is present on the windward liquid jet surface, the primary atomization of the liquid jet can also occur through the formation of the Kelvin-Helmholtz instability (KHI) \cite{xiao2016large,behzad2016surface,medipati2025elliptic} on the lateral liquid jet surfaces, resulting from the intense streamwise shear between the liquid jet and the high-speed air\citep{xiao2016large,chai2015numerical}. This resembles the atomization of a single droplet in supersonic flow \citep{joseph1999breakup}. The formation of KHI on the lateral surfaces leads to the formation of liquid tongues that form perpendicular to the lateral surface, and these will undergo further atomization through RTI due to the high-speed crossflow, leading to vigorous mass stripping, as discussed by \citep{medipati2025elliptic,varga2003initial}.  As $J$ increases in the present study (for cases $AR$ = 0.3 and 1), this cascade of atomization (KHI$\rightarrow$RTI) and the mass stripping from the lateral surfaces occurs at higher transverse heights, seen as less darker regions on the leeward side of the jet for the higher $J$ cases (Fig. \ref{fig:4}(ii)) in contrast to the lower cases $J$ (Fig. \ref{fig:4}(i)). To further confirm this and to understand the influence of $J$ on the morphology of the spray plume due to the difference in mass stripping behavior, planar laser Mie scattering (PLMS) is employed using a laser sheet perpendicular to the free-stream direction at a given streamwise position ($x/D$) for different $J$ and for each orifice $AR$ case. A large number of instantaneous images of about 500 are captured for each $J$ case, and an ensemble-averaged image is obtained whose variation with $J$ at $x/D$ = 60 and for orifice $AR$ = 1 is shown in figure~\ref{fig:5}. It is seen that, in the case of lower $J$ (Fig. ~\ref{fig:5}(i)), since the mass stripping from the lateral surfaces are taking place at lower transverse heights, the averaged spray width in the span-wise direction ($z/D$) near the tunnel floor ($y/D$ = 0) is significantly higher and is about 13 $D$. In contrast, in the case of higher $J$ (Fig. \ref{fig:5}(ii)), the near-floor spray width is negligibly small (3 $D$), indicating the significant reduction in mass stripping at lower transverse heights. This leads to significant changes in the shape of the overall spray plume width and cross-section, resulting in an inverted 'U' and an oval-shaped trend, in the cases of lower and higher $J$, respectively. These disparities with $J$ are most likely due to the formation of a stronger bow shock wave at higher $J$ cases, resulting in a lower free-stream velocity after the shock, which leads to a significant decrease in the shear acting on the liquid jet lateral surfaces at lower transverse heights. This can ultimately lead to a significant difference in the distribution of droplets across the spray plume due to the variation in $J$. Whereas in the case of $AR$ = 3.3, the frontal area is significantly higher, resulting in the intense acceleration of the liquid jet, and the primary atomization of the liquid jet is purely through RTI, leading to catastrophic breakup of the liquid jet, and is invariant with $J$.
 
 As shown, $J$ has a significant influence on the formation of windward surface waves of the liquid jet and their variation with time for different orifice $AR$ cases. These variations in the windward surface waves with $J$ will have immediate consequences on the shock waves formed and the overall unsteady behavior of the liquid jet and shock wave, and are quantitatively discussed in the sections~\ref{Shock structures} and \ref{Unsteady behavior of penetration height and shock position}.

\begin{figure*}
        \centering
        \includegraphics[width = 0.9\textwidth]{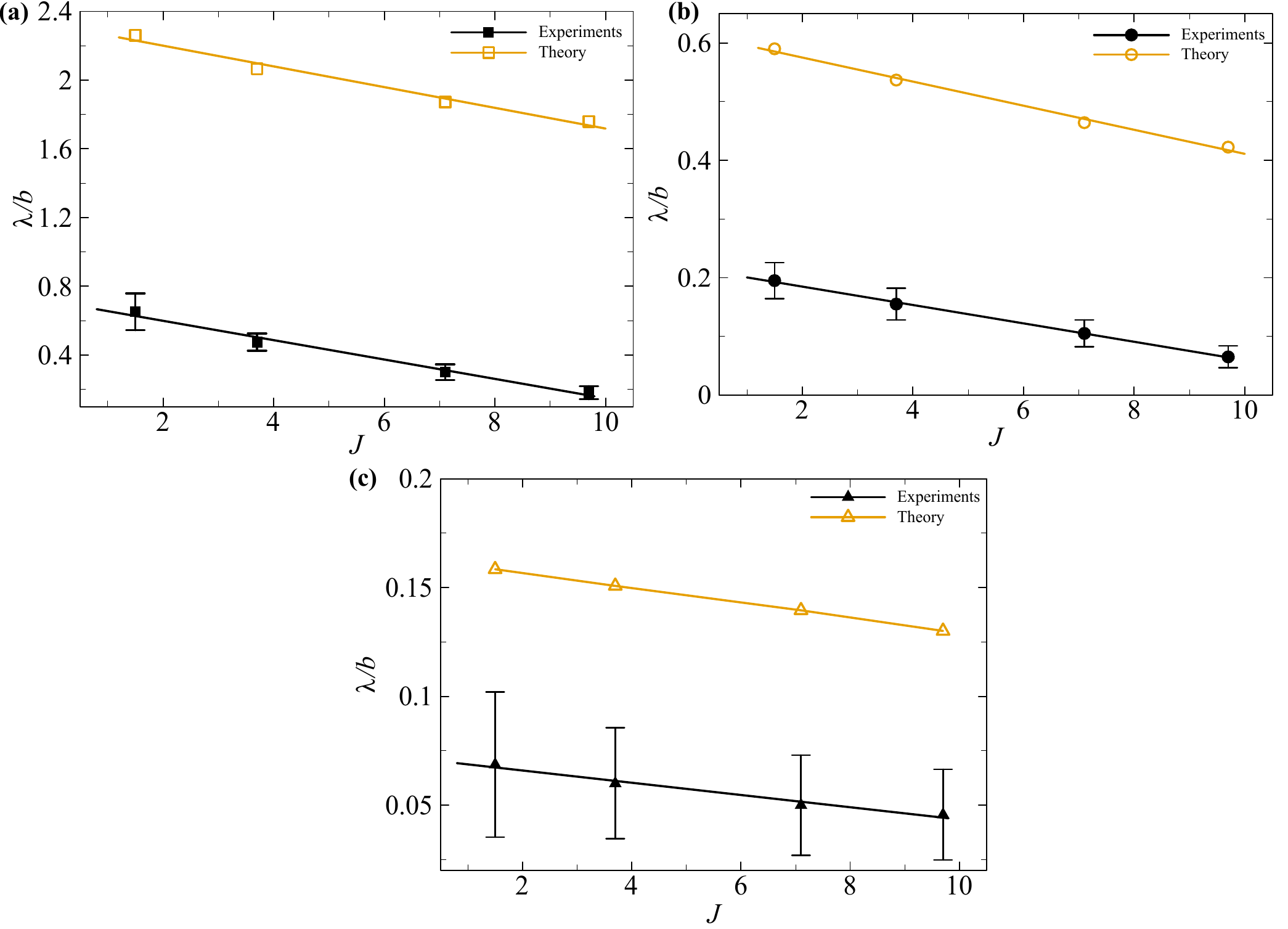}
        \caption{Variation of measured dimensionless surface wavelength with $J$ for (a) $AR$ = 0.3, (b) $AR$ = 1, and (c) $AR$ = 3.3. The experimental and theoretical values of the most unstable RT wavelengths and their variation with $J$ are shown in filled black and unfilled orange symbols, respectively.}
        \label{fig:6}
\end{figure*}

  We now proceed to present the most amplified surface wavelength values determined from experimental visualizations for each of the different $J$ cases and separately for each $AR$ case, and then compare them with theoretical calculations. Utilizing the instantaneous visualizations as shown in figure~\ref{fig:4} for different $J$ and for each $AR$, the distance between the two troughs on the windward surface of the jet is measured and is denoted as surface wavelength ($\lambda$) in the corresponding left side zoomed-in images.  For each experimental condition, a large set of about 200 instantaneous images were used, and an average wavelength was determined.  This surface wavelength is normalized with the corresponding frontal dimension of the orifice ($b$), and its variation with $J$ for (a) $AR$ = 0.3, (b) $AR$ = 1, and (c) $AR$ = 3.3, as shown in figure~\ref{fig:6}. As observed in figure~\ref{fig:6}, irrespective of the $AR$, as the $J$ increases, the value of $\lambda/b$ decreases roughly linearly. For example, in the case of $AR$ = 1 (figure \ref{fig:6}b), the value of $\lambda/b$ has decreased from 0.2 to 0.06 as $J$ increases from 1.5 to 9.7. This rapid decrease in $\lambda/b$ is due to the increase in the acceleration of the liquid jet as a result of higher drag forces on the liquid jet, as discussed earlier. A similar trend in $\lambda/b$ with $J$ is also found for $AR$ = 0.3 and $AR$ = 3.3 cases as depicted in figure~\ref{fig:6}a and figure~\ref{fig:6}c, respectively. It may be noted that similar variation in $\lambda/b$ with $J$ is also seen in the case of the recent experimental study of circular liquid jets in subsonic cross-flow \cite{shaw2024breakup}. The influence of $AR$ on the value of $\lambda/b$ is also observed while the comparison is drawn for a fixed $J$. For the $J$ = 1.5 case, when the liquid jet is injected with $AR$ = 0.3, the value of $\lambda/b$ (0.68) is significantly larger than the other cases due to the lower acceleration of the liquid jet, owing to the lower drag experienced by the liquid jet (lower frontal area).  In contrast, in the case of $AR$ = 3.3, the value of $\lambda/b$ (0.07) is small because of the rapid accelerations experienced by the liquid jet, due to the higher drag experienced by the liquid jet (higher frontal area), as mentioned earlier.      

 Having seen the experimentally determined variations with $J$, we shall now see the calculations of the theoretically obtained values of the most unstable RT wavelength for different $J$ cases and for each orifice $AR$ case. The most unstable wavelength pertaining to maximum growth rate that is responsible for the transverse destabilization of the liquid jet due to the acceleration ($a$) of the low-density air over a high-density liquid ($\rho_l$) is theoretically expressed as \citep{chandrasekhar2013hydrodynamic, joseph1999breakup,varkevisser2024effect,medipati2025elliptic}:
 \begin{equation}
    \label{eq:surface wavelength}
    \lambda = 2\pi \sqrt{\frac{3\sigma}{\rho_l a}}.
\end{equation}
This clearly shows that the most unstable wavelength is a strong function of the acceleration ($a$) of the given liquid jet by the surrounding gas.  We now see how to modify this equation and estimate the value of $a$ for different $J$ cases and for a given $AR$. 

\begin{figure*}
        \centering
        \includegraphics[width = 0.4\textwidth]{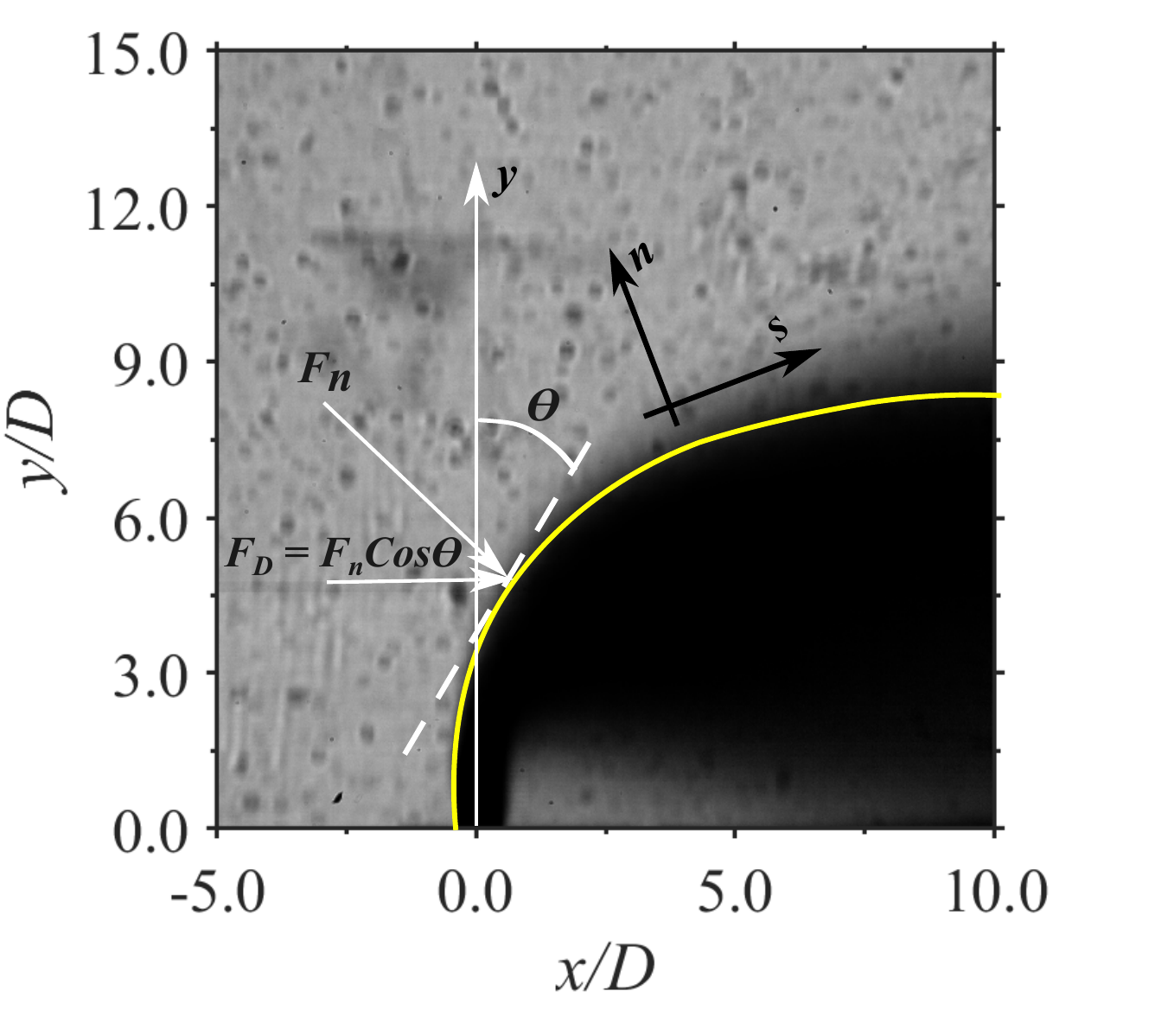}
        \caption{Ensemble-averaged image of the circular jet for $J$ = 1.5 case obtained by averaging the large set of instantaneous images acquired using high-speed shadowgraphy. The yellow-colored curve indicates the jet penetration trajectory.  The coordinates parallel and normal to the windward jet surface are $s$ and $n$, respectively.  $\theta$, $F_n$, and $F_D$ represent the windward jet deflection angle, aerodynamic force acting normal to the windward jet surface, and the drag force acting in the streamwise direction, respectively.  The dashed line corresponds to the tangent drawn to the jet trajectory.}
        \label{fig:7}
\end{figure*}

As discussed earlier, the value of $a$ is a strong function of the drag forces acting on the windward surface, which in turn varies with $J$ in the present study for a given $AR$ case.  To establish a relation between them and compute the value of $a$, we show in figure~\ref{fig:7}, an ensemble-averaged image along with the mean penetration trajectory (yellow colored curve) and the aerodynamic forces acting normal to the windward jet surface by the high-speed air ($F_n$) \cite{mashayek2008improved,broumand2016two,xie2025numerical}. Because of the drag force ($F_D$) exerted by the high-speed crossflow, the liquid jet accelerates in the streamwise direction and deflects into the cross-flow. The windward jet deflection angle ($\theta$) is measured at a transverse location where the windward jet mean penetration trajectory crosses the vertical axis passing through the orifice center ($y$).  Considering a volume of liquid ($V_l$) that is accelerating and turning into the streamwise direction due to the presence of air, the streamwise force balance of the liquid volume results in an acceleration \citep{joseph1999breakup} given by:
\begin{equation}
    \label{eq:accelaration}
    a = \frac{F_D}{m_l} = \frac{0.5C_D \rho_{2} U_{2}^2 A_f \cos^3 \theta}{\rho_l V_l}, 
\end{equation} 
 where $\rho_2$ and $U_2$ indicate density and velocity of the free stream air behind the normal shock (corresponding to $M_\infty$ = 2.5) \citep{medipati2025elliptic} and  $C_D$, $\rho_l$, and $A_f$ represent the coefficient of drag, liquid density, and frontal/exposed area of the liquid jet to the high-speed air.  Equation~\ref{eq:accelaration} strongly emphasizes the fact that for a given cross-flow condition, fixed orifice $AR$ (fixed $A_f$) and $V_l$, the acceleration of the liquid jet is purely a function of the Cosine of the jet deflection angle ($\theta$), which in turn is a function of $J$ in the present study.  It is important to note that in the present study, an increase in $J$ (higher penetration) leads to lower $\theta$, resulting in a significant increase in the acceleration of the liquid jet. Hence, for a given $AR$, when the liquid jets are injected with higher momentum into the cross-flow, they exhibit RT waves that are smaller in wavelength because of the enhanced accelerations experienced by the liquid jet.  It may be noted that the calculated value of $a$ in the present study is found to be of the order of $10^4$ times the acceleration due to gravity ($g$).  This is consistent with the calculations of \cite{joseph1999breakup} for a single droplet in a supersonic flow study.   
Substituting eqn.~\ref{eq:accelaration} for $a$ into eqn.~\ref{eq:surface wavelength} and by replacing $V_l$ as a product of $A_c$ (orifice cross-section area) and a unit thickness, and $A_f$ as a product of $b$ (frontal dimension of the orifice) and a unit thickness, yields the equation for the normalized most unstable wavelength ($\lambda/b$) as a function of $C_D$, $We_\mathrm{eff}$, and $\cos \theta$ as
\begin{equation}
    \label{eq:lambda by b}
    \frac{\lambda}{b} = \frac{2\pi}{b}\sqrt{\frac{6 A_c}{C_D We_{eff} \cos^3 \theta}},
\end{equation}
where $We_\mathrm{eff}$ represents the effective cross-flow Weber number ($ =  \rho_2 U_2^2 b/\sigma$), which is defined based on the properties behind the normal shock wave, and normalized by the frontal dimension of the orifice ($b$) \cite{xiao2016large,medipati2025elliptic}. It may be noted that for a given orifice $AR$, $We_\mathrm{eff}$ is fixed and an increase in orifice $AR$ leads to an increase in $We_\mathrm{eff}$.  

Finally, to theoretically estimate the value of $\lambda/b$ from eqn.~\ref{eq:lambda by b}, both $C_D$ and $\cos \theta$ values are needed.  The drag coefficient, $C_D$, which is a function of orifice aspect ratio for a given cross-flow condition is taken from the supersonic flow over solid bodies of different aspect ratio studies\cite{gowen1953drag,heddleson1957summary}, whose values are shown in Table~\ref{tab:drag coefficient}. For a given $AR$ ($We_\mathrm{eff}$), the values of $\theta$ for different $J$ are determined from the averaged images, as shown in figure~\ref{fig:7}.

\begin{table}[ht]
\centering
  \begin{tabular}{  c   c   c   c   }
  \hline
 $AR$ &  $A_f(mm^2)$  & $ C_D$ &  $V_l(mm^3)$   \\
 \hline
  0.3  &  0.61  & 0.6 & 1.06  \\
  1 & 1.2  & 1.5   & 1.1 \\
  3.3 & 2.23 & 1.8  & 1.06 \\
  \hline
 \end{tabular}
\caption{Values of $A_f$, $C_D$, and $V_l$ used in the present model for different values of $AR$.}
\label{tab:drag coefficient}
\end{table}

The predicted most amplified $\lambda/b$ from eqn.~\ref{eq:lambda by b} and its variation with $J$ are reported and marked as unfilled orange symbols in figures~\ref{fig:6} (a), (b), and (c) for $AR$ = 0.3, 1, and 3.3, respectively.  Both the experimental and theoretical estimations capture a similar decreasing trend with $J$. From these measurements and the analysis, it can be concluded that as $J$ increases, the value of $\lambda/b$ rapidly decreases, confirming that the formation of the RT waves are smaller in wavelength due to the increase in the acceleration of the liquid jet caused by the higher drag of the high-speed air, resulting from the lower jet deflection into cross-flow ($\theta$).  These increased accelerations of the liquid jet, resulting from an increase in $J$, will help the jet undergo early breakup, which can eventually lead to the formation of fine droplets. On the other hand, for a given $J$, as the orifice $AR$ increases from 0.3 to 3.3, the value of $C_D$ (0.6 - 1.8) and $We_\mathrm{eff}$ (440 to 1600) increases significantly, resulting in a rapid decrease in $\lambda/b$ with $AR$ as well.  This also results in the formation of smaller RT waves due to the intense acceleration of the liquid jet caused by the high-speed air in the case of $AR$ = 3.3, due to the larger drag forces, resulting in the eventual formation of finer droplets\citep{medipati2025elliptic}. 
\begin{figure*}
        \centering
        \includegraphics[width = 0.9\textwidth]{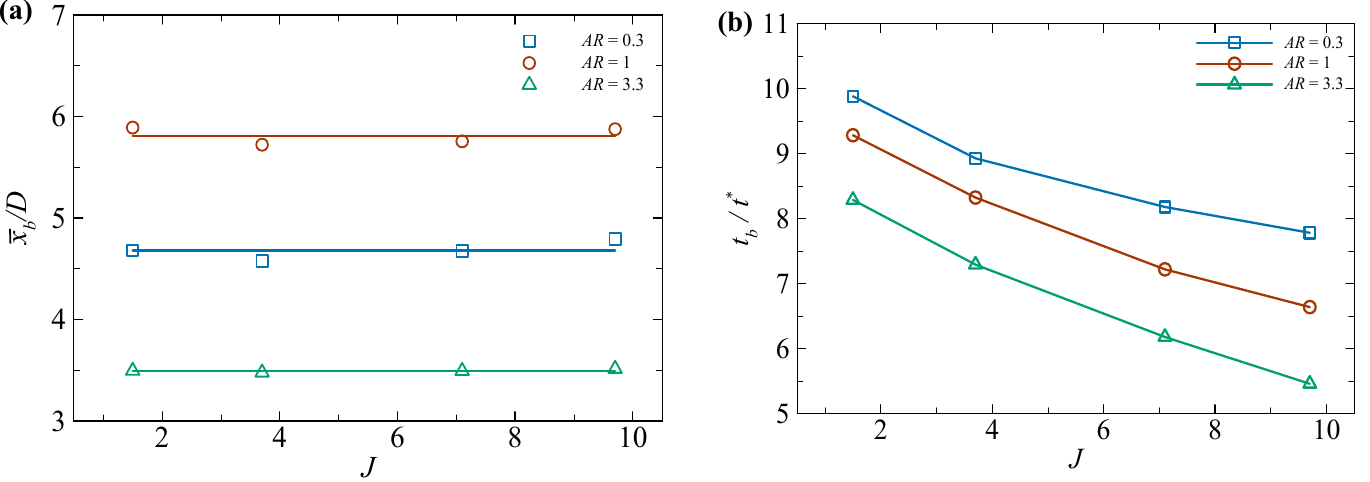}
        \caption{(a) Variation of mean streamwise breakup location with $J$, and (b) Variation of normalized breakup time of the liquid column with $J$.}
        \label{fig:8}
\end{figure*}

It is also evident from figure~\ref{fig:4} that the initial RTI waves ($\lambda/b$) formed on the windward jet surface grow rapidly as the jet penetrates into the cross-flow due to the continuous acceleration of the liquid jet by the high-speed air \cite{varkevisser2024effect}.  This eventually results in the jet column fracture, as marked with a red dot in figure~\ref{fig:4}b.  For each experimental condition, the instantaneous breakup location from the orifice center, in both the streamwise ($x_b$) and cross-streamwise ($y_b$) directions, is identified and measured. Using a large number of instantaneous values, its time-averaged values ($\overline{x_b}$/\textit{D} and $\overline{y_b}$/\textit{D}) are calculated.  It is found that the value of $\overline{x_b}$/\textit{D} is invariant with $J$, for a given $AR$, as shown in figure \ref{fig:8}a, which is in close agreement with the liquid jet in subsonic cross-flow studies \cite{broumand2016liquid,sallam2004breakup,wu1997breakup}.  However, a strong dependency is observed with $AR$ as discussed in our previous study\cite{medipati2023liquid}. As expected, with an increase in $J$, the value of $\overline{y_b}$/\textit{D} increases due to the significant increase in jet penetration height \cite{broumand2016liquid,sallam2004breakup,wu1997breakup,medipati2025elliptic}. To understand the impact of increased accelerations of the liquid jet due to an increase in $J$, the jet breakup time, \begin{math}t_b = \sqrt{\frac{2 \overline{x_b}}{a}}\end{math} for different $J$ and for each orifice $AR$ case is calculated similarly to the droplet in supersonic flow \citep{engel1958fragmentation,nicholls1969aerodynamic,hsiang1992near}, liquid jet in subsonic\citep{sallam2004breakup,wu1997breakup,broumand2016liquid,varkevisser2024effect} and supersonic cross-flow studies \citep{medipati2025elliptic}. Figure \ref{fig:8}b shows the variation of $t_b$ with $J$ for different $AR$.  The value of $t_b$ is normalized with the corresponding inertial time scale of the cross-flow, \begin{math}t^* = {\frac{b}{U_\infty}}\sqrt{\frac{\rho_j}{\rho_\infty}}\end{math} \citep{hsiang1992near,broumand2016liquid}. For a given $AR$, as the value of $\overline{x_b}$ is a constant irrespective of different $J$ (figure \ref{fig:8}a), hence, the value of $t_b$ is inversely related to the $\sqrt{a}$ of the liquid jet, whose value significantly increases with an increase in $J$ (from eqn.~\ref{eq:accelaration}).  This results in a significant decrease in the normalized liquid breakup time ($t_b/t^*$) with an increase in $J$, as shown in figure~\ref{fig:8}b.  This implies there is enhancement in the atomization of the liquid jet as $J$ increases, due to the increased accelerations experienced by the liquid jet, as discussed earlier.  A similar decreasing trend of $t_b/t^*$ with $J$ is also seen in the subsonic cross-flow studies by \cite{lubarsky2012experimental,shaw2024breakup}. In addition to that, for a given $J$, as the $AR$ increases, the value of $t_b/t^*$ also decreases further, indicating the further enhancement in the atomization due to the dominant RTI in the case of $AR$ = 3.3, which is in line with \citep{medipati2025elliptic}.  Combining the $\lambda/b$ and $t_b/t^*$ calculations, it can be stated that as the liquid jet exhibits smaller $\lambda/b$ values  (enhanced accelerations of the liquid jet), then $t_b/t^*$ becomes lower (early breakup), and hence, the droplets downstream will be smaller.  Also, note that the value of $t_b/t^*$ found from the present and previous studies \citep{medipati2025elliptic} are in close agreement with the atomization of droplets in supersonic cross-flow studies \citep{engel1958fragmentation,nicholls1969aerodynamic,hsiang1992near}. This suggests that the atomization mechanism of a (single) droplet in supersonic flow has similarities with that of a liquid jet in supersonic cross-flow.  
\subsection{Shock structures}
\label{Shock structures}

Having seen the influence of $J$ on the formation of windward surface waves, the initial atomization mechanism, and break time scales of the liquid jet, now we shall present and discuss the influence of $J$ on the formation of the complex shock structures and its strength upstream of the liquid jet, which can introduce a complex flow field around the liquid jet during the atomization process.  Finally, how these flow-field interactions are altered due to the variations in $AR$ is also discussed. 
\begin{figure*}
        \centering
        \includegraphics[width = 0.9\textwidth]{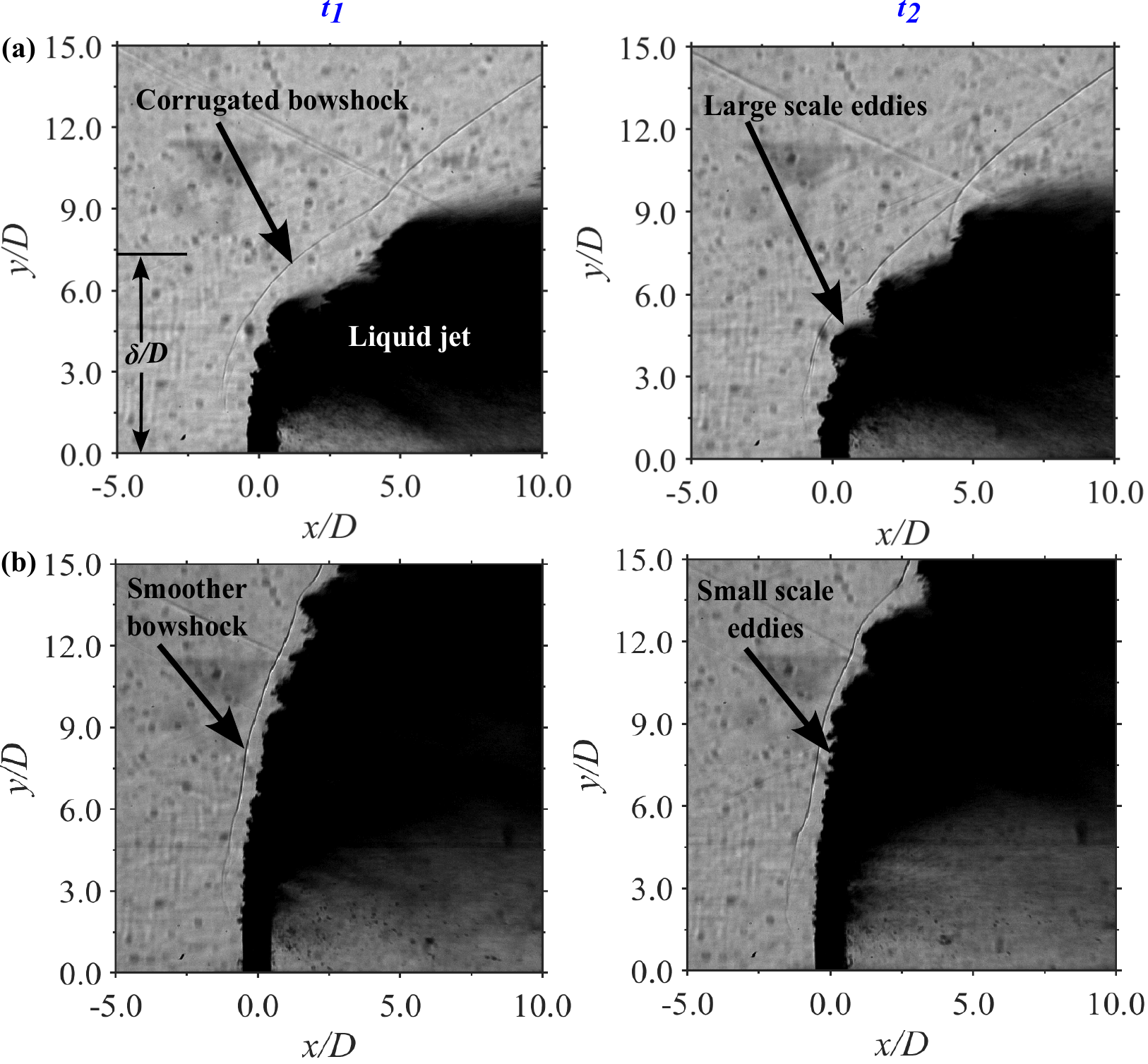}
        \caption{Instantaneous visualizations captured using high-speed shadowgraphy highlighting the shock structures, the windward spray edge, and the leeward side wake region for (a) $J$ = 1.5, and (b) $J$ = 9.7 and for $AR$ = 1. The left and right visualizations are acquired at different instants ($t_1$ and $t_2$).}
        \label{fig:9}
\end{figure*}
\begin{figure*}
        \centering
        \includegraphics[width = 0.9\textwidth]{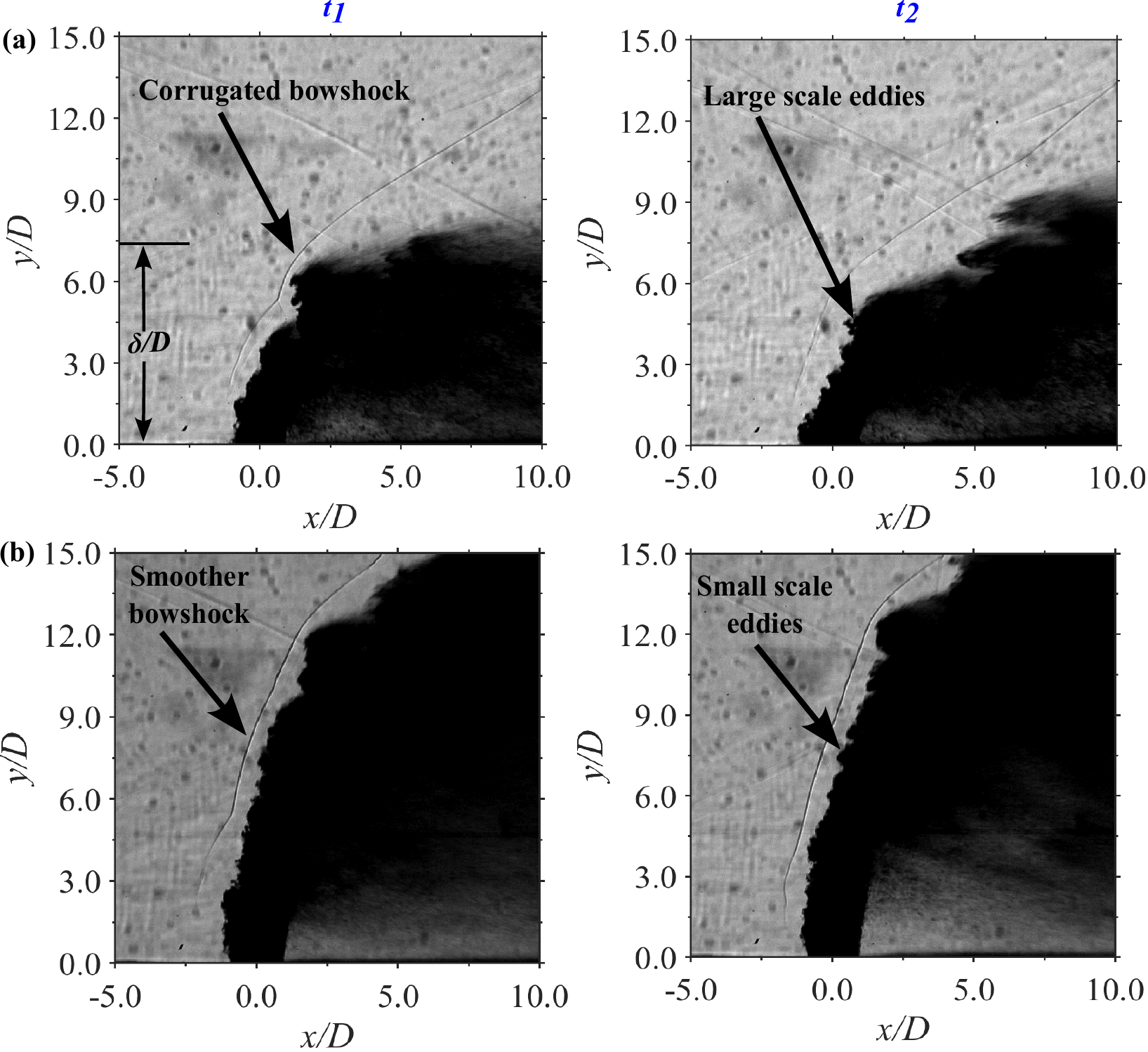}
        \caption{Instantaneous visualizations captured using high-speed shadowgraphy highlighting the shock structures, the windward spray edge, and the leeward side wake region for (a) $J$ = 1.5, and (b) $J$ = 9.7 and for $AR$ = 0.3. The left and right visualizations are acquired at different instants ($t_1$ and $t_2$).}
        \label{fig:10}
\end{figure*}
\begin{figure*}
        \centering
        \includegraphics[width = 0.9\textwidth]{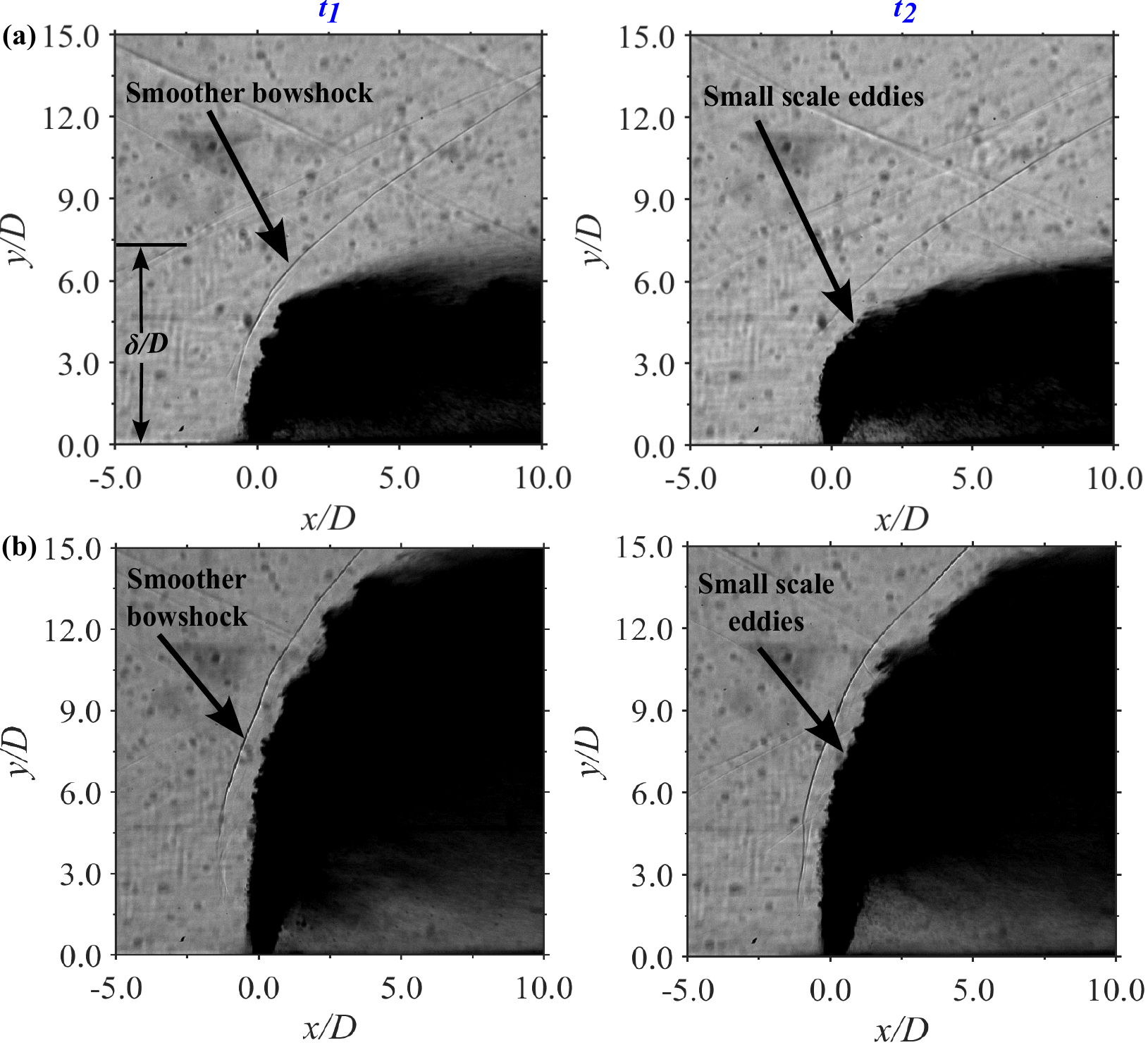}
        \caption{Instantaneous visualizations captured using high-speed shadowgraphy highlighting the shock structures, the windward spray edge, and the leeward side wake region for (a) $J$ = 1.5, and (b) $J$ = 9.7 and for $AR$ = 3.3. The left and right visualizations are acquired at different instants ($t_1$ and $t_2$).}
        \label{fig:11}
\end{figure*}
\begin{figure*}
        \centering
        \includegraphics[width = 0.9\textwidth]{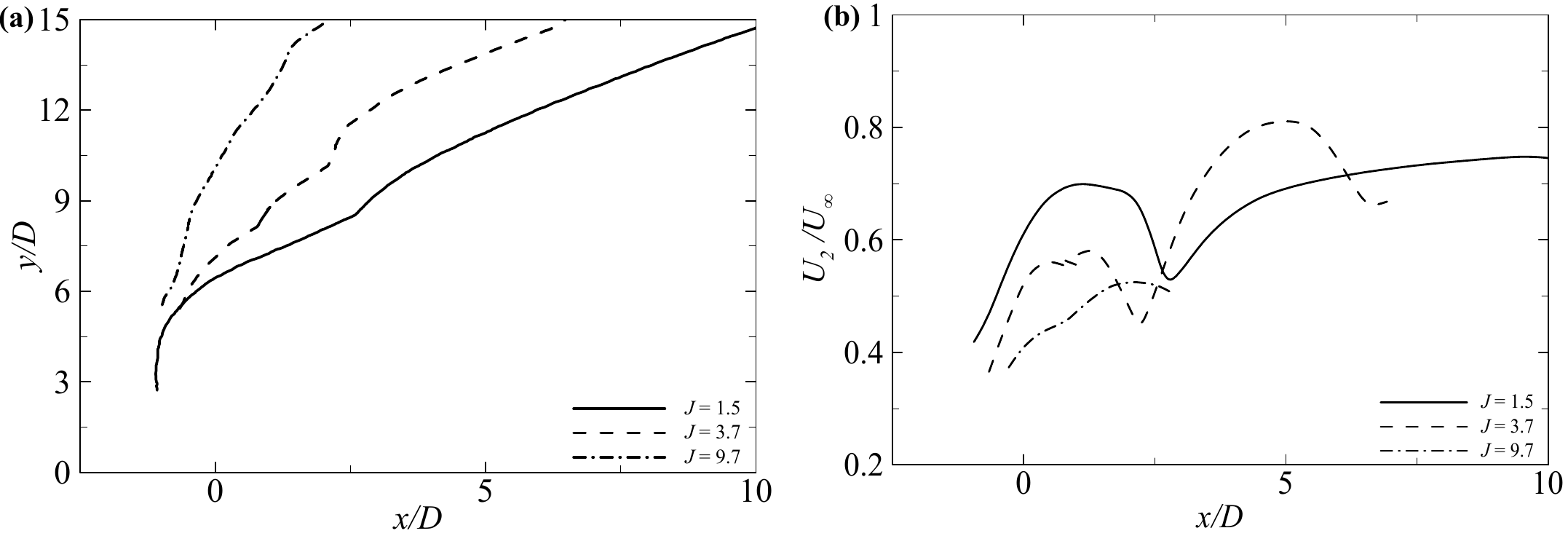}
        \caption{Influence of $J$ on the instantaneous (a) shock traces, and (b) variation of the shock-induced velocity along the corresponding shock traces with $x/D$ to highlight the formation of corrugations on the velocity variation around the liquid structure for $J$ = 1.5, 3.7, and 9.7, and $AR$ = 1. The shock-induced velocities are normalized with the free-stream velocity.}
        \label{fig:12}
\end{figure*}

Figures~\ref{fig:9}a \& \ref{fig:9}b shows two instantaneous visualizations for $J$ = 1.5, and $J$ = 9.7 for $AR$ = 1, respectively. These images are captured using high-speed shadowgraphy by zooming a narrow field of view of about 15$D$ x 15$D$ near the jet exit.  The value of mean boundary layer thickness $\delta/D$ (measured using PIV in the absence of the jet) is also indicated to emphasize the extent of the jet penetration height into the cross-flow with respect to  $\delta/D$.  These instantaneous visualizations help us understand the influence of $J$ on the formation of shock structures ahead of the injected liquid jet and their unsteady interaction with the windward spray edge of the liquid jet and with the oncoming boundary layer structures.    

As the jet penetrates into the supersonic cross-flow, it acts as an obstruction to the oncoming supersonic flow, resulting in the formation of a 3D bowshock wave \citep{karagozian2010transverse,lin2004structures,medipati2023liquid,medipati2025elliptic} (a thin dark curved line) in the upstream of the liquid jet (right-side dark region of the image). It is found that jets injected with higher momentum flux ratios ($J$ = 9.7) lead to the formation of stronger oblique shock waves by displacing to a more upstream position due to higher jet penetration into the cross-flow.  In the case of $AR$ = 1, as the $J$ increases from 1.5 to 9.7, the mean jet penetration height (at $x$ = 1$D$) increases from 5$D$ to 12$D$, resulting in the upstream displacement of the mean shock location in the streamwise direction from 7$D$ to 13$D$ along with an increase in shock angle from about 45$\degree$ to 75$\degree$. A similar variation with $J$ is also observed for $AR$ = 0.3, and 3.3 cases as shown in Figures \ref{fig:10}a \& \ref{fig:10}b and Figures \ref{fig:11}a \& \ref{fig:11}b, for (a) $J$ = 1.5 and (b) $J$ = 9.7, respectively. This indicates that $J$ plays a crucial role in controlling the shock strength and location, in addition to the jet penetration height.                 

As discussed earlier (section~\ref{Atomization mechanism}), in the case of lower $J$ (= 1.5) and $AR$ = 1 (figure \ref{fig:9}a), the surface waves formed on the windward side of the liquid jet are larger due to the reduced acceleration of the jet, due to the reduced drag force.  This results in the formation of a corrugated bow shock wave ahead of the jet with large variations in the local curvature/strength of the shock wave.  This leads to large variations in the flow field as experienced by the liquid jet.  In contrast, as the $J$ increases gradually to 9.7 (figure \ref{fig:9}b), the drag forces acting on the windward jet surface have gradually increased, resulting in the gradual increase in the acceleration of the liquid jet with the formation of smaller RT waves.  Hence, the bow shock waves formed are relatively smoother in these cases.  Further, in the case of lower $J$ (figure \ref{fig:9}a), as mentioned earlier, an irregular formation of the windward surface waves between instants is due to stronger interaction with the incoming boundary layer streaks, which leads to the formation of a complex corrugated shock structure with large variations between instants.  This can be seen clearly by comparing the images acquired during a single experiment of $J$ = 1.5 case at two different instants ($t_1$ and $t_2$).  These are shown in the left and right side visualizations of figures~\ref{fig:9}a ($AR$ = 1), figures \ref{fig:10}a ($AR$ = 0.3), and figures \ref{fig:11}a ($AR$ = 3.3). For instance, at $t_1$ in figure \ref{fig:9}a, the shock is pushed more downstream due to the lower penetration height of the liquid jet and is in more contact with the tunnel floor in comparison with the $t_2$ instant.  Additionally, at $y/D$ = 9, the liquid jet undergoes large-scale deformation, resulting in the local upstream displacement of the shock with a higher strength, which forms a bumpy shock.  These dynamics have gradually reduced at $J$ of 9.7 (figures~\ref{fig:9}b, \ref{fig:10}b, and \ref{fig:11}b). 
This exemplifies that due to the irregular and unsteady formation of windward surface waves/liquid structures in the lower momentum liquid jets ($J$ = 1.5), the formation of corrugations on the shock wave between instants has increased significantly. 

A similar decrease in the interaction dynamics with $J$ is also observed for other orifice $AR$ cases, but the scale of interaction dynamics is seen to have a significant influence on $AR$.  Irrespective of the $J$, in the case of $AR$ = 0.3 (figures \ref{fig:10}a \& \ref{fig:10}b), and 3.3 (figures \ref{fig:11}a \& \ref{fig:11}b), the value of $\lambda/b$ is larger and smaller (figure \ref{fig:6}) due to the lower and higher acceleration of the liquid jet, hence the shock waves are more corrugated and smoother, respectively, and is consistent with our previous study \citep{medipati2025elliptic}.

The influence of $J$ on the variations in shock structures as seen in the visualizations can be understood by plotting sample instantaneous shock positions for different $J$, as depicted in figure~\ref{fig:12}a.  These are plotted for a given orifice aspect ratio ($AR$ = 1).  As seen from the visualizations, as $J$ increases, the instantaneous shocks formed are relatively more oblique and are displaced to more upstream locations. In addition to that, a large increase in the curvature of the instantaneous traces at lower $J$ ( = 1.5) highlights the increased corrugations of the shock wave.  The corrugations formed on the shock wave are further analyzed by computing the corresponding post-shock velocity of free stream air ($U_{2}$) variation along the streamwise direction, as shown in figure~\ref{fig:12}b.  These calculations are performed using the inviscid shock relations similar to \citep{ben2006time,medipati2025elliptic}. From figure~\ref{fig:12}b, it is evident that as $J$ increases, the value of $U_2$ has reduced significantly due to the increase in shock strength. For example, at $x/D$ = 0, the value of $U_2$ has decreased from 0.6$U_\infty$ to 0.4$U_\infty$, due to an increase in  $J$ from 1.5 to 9.7. This can lead to lower shear on the lateral surfaces at lower jet transverse heights in cases of higher $J$, resulting in lower mass stripping from the jet's lateral surfaces through KHI$\rightarrow$RTI, as discussed earlier. Thus, the variation in shock strength with $J$ can impact the amount of mass being stripped from the jet surface, which can lead to a significant influence on the distribution of droplets across the plume. Furthermore, it is also observed that in the case of $J$ = 1.5, along the shock wave, the instantaneous value of $U_2$ fluctuates between 0.4$U_\infty$ and 0.8$U_\infty$.  However, the fluctuations have significantly reduced as the $J$ gradually increases to 9.7.  Moreover, in the lower $J$ case, the large instantaneous variation in the shock corrugations between the instants (figure \ref{fig:9}a) will lead to fluctuations in $U_2$ with time. This leads to large variations in the velocity field experienced by the liquid jet in the case of lower $J$, which can eventually result in the large-scale unsteadiness in the downstream spray characteristics.  Hence, the instantaneous shock curvature and its position are linked to the occurrence of irregular liquid structures on the windward side of the liquid jet and are more pronounced in the case of $J$ = 1.5 due to the presence of higher $\lambda/b$ and its variation with time due to the increased interaction with the oncoming boundary layer fluctuations.  The formation of irregular liquid structures and their influence on the development of the corrugations on the shock wave are further intensified in the case of $AR$ = 0.3 (figure \ref{fig:10}) due to the notably higher $\lambda/b$ values, hence the corresponding variations in the post-shock velocity are much more significant. Whereas in the case of $AR$ = 3.3 (figure \ref{fig:11}), the RT waves are very small (larger accelerations), hence the shock waves are free from corrugations with a smoother post-shock velocity variation, as discussed in detail in our previous study \cite{medipati2025elliptic}. 

\section{Unsteady behavior of penetration height and shock position}
\label{Unsteady behavior of penetration height and shock position}

The instantaneous visualizations as shown in the left and right side figures of \ref{fig:9} (a \& b), \ref{fig:10} (a \& b), and \ref{fig:11} (a \& b), elucidate the influence of $J$ for each $AR$ on the unsteadiness in jet penetration height and shock motion.  We shall now quantify this unsteadiness and show how $J$ affects the unsteadiness due to the differences in the interaction with the oncoming boundary layer streaks.  Three cases of $J$ are chosen for representation purposes in each $AR$ case, such that $J$ = 1.5, 3.7, and 9.7 indicate low, moderate, and high $J$ cases, respectively.  This is based on the differences in the formation of the RT waves, shock structures, and the comparison of mean jet penetration height ($\overline{h}$/$D$) with respect to the mean incoming boundary thickness ($\delta$/$D$).  Finally, the variations in the orifice $AR$ on the unsteadiness are also discussed. 
\begin{figure*}
        \centering
        \includegraphics[width = 0.9\textwidth]{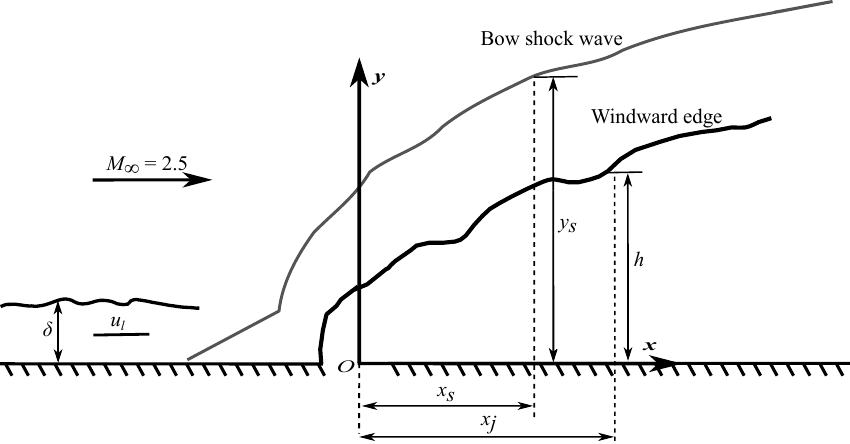}
        \caption{Schematic highlighting the instantaneous flow features of the liquid jet in supersonic flow.  $u_l$ is the line-averaged velocity determined from PIV measurements,  $x_s$, $y_s$, and $x_j$, $h$ indicate the instantaneous shock and jet locations in the streamwise and transverse directions, respectively.}
        \label{fig:13}
\end{figure*}

\begin{figure*}
        \centering
        \includegraphics[width = 0.9\textwidth]{Figures/Figure14.pdf}
        \caption{Instantaneous and mean traces of bow shock wave (left) and windward spray edge (right) of liquid jet for (a) $J$ = 1.5, (b) $J$ = 3.7, and (c) $J$ = 9.7, and $AR$ = 1. In all the plots, instantaneous traces of the shock wave and the windward spray edge are shown in blue.  The corresponding mean traces are shown in thick black line curves.  The horizontal pink colored line marked at a cross-streamwise location ($y/D$ = 8) represents the width of the fluctuations.}
        \label{fig:14}
\end{figure*}
\begin{figure*}
        \centering
        \includegraphics[width = 0.9\textwidth]{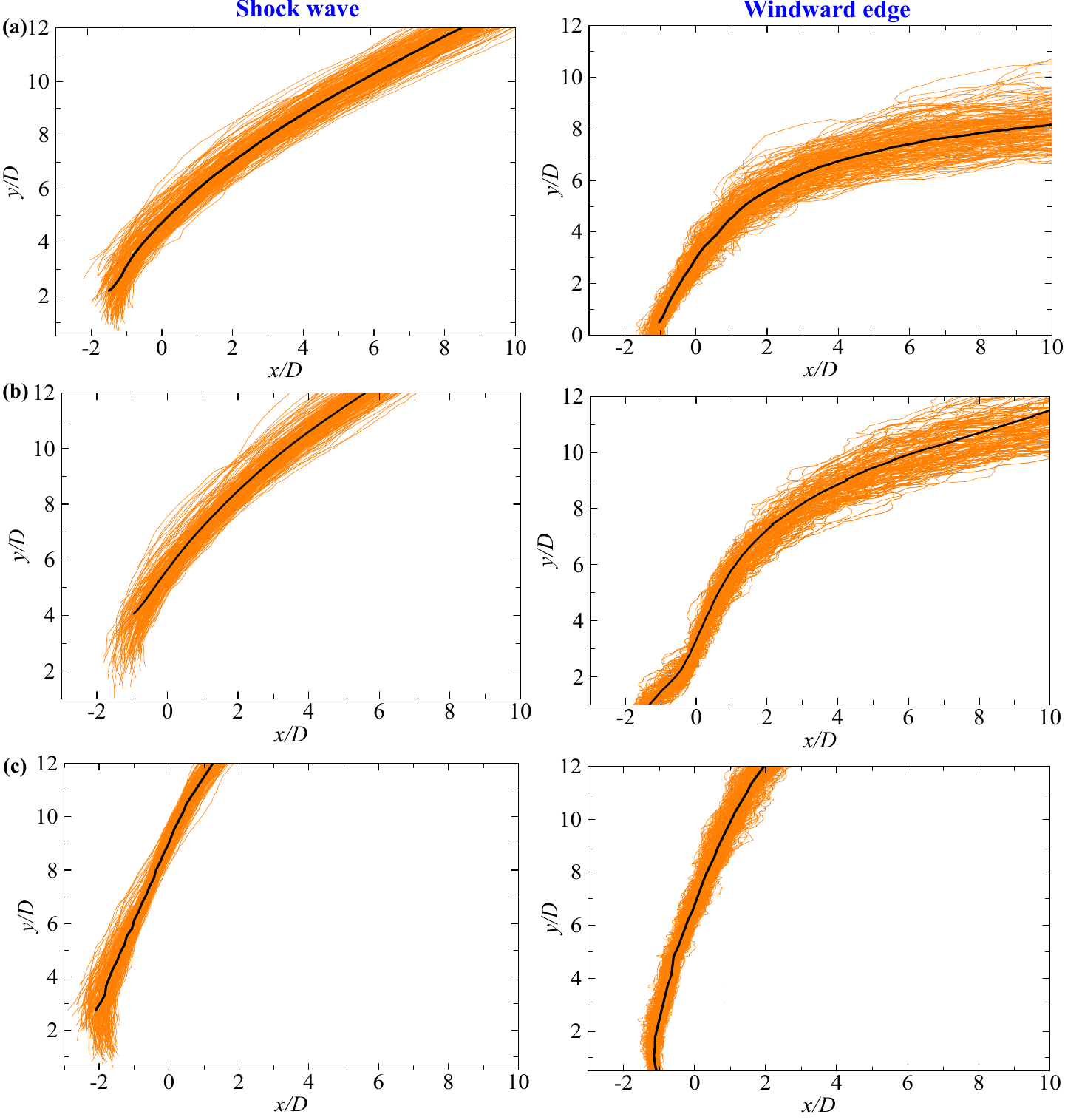}
        \caption{Instantaneous and mean traces of bow shock wave (left) and windward spray edge (right) of liquid jet for (a) $J$ = 1.5, (b) $J$ = 3.7, and (c) $J$ = 9.7, and $AR$ = 0.3.  In all the plots, instantaneous traces of the shock wave and the windward spray edge are shown in orange. The corresponding mean traces are shown in thick black line curves.}
        \label{fig:15}
\end{figure*}
\begin{figure*}
        \centering
        \includegraphics[width = 0.9\textwidth]{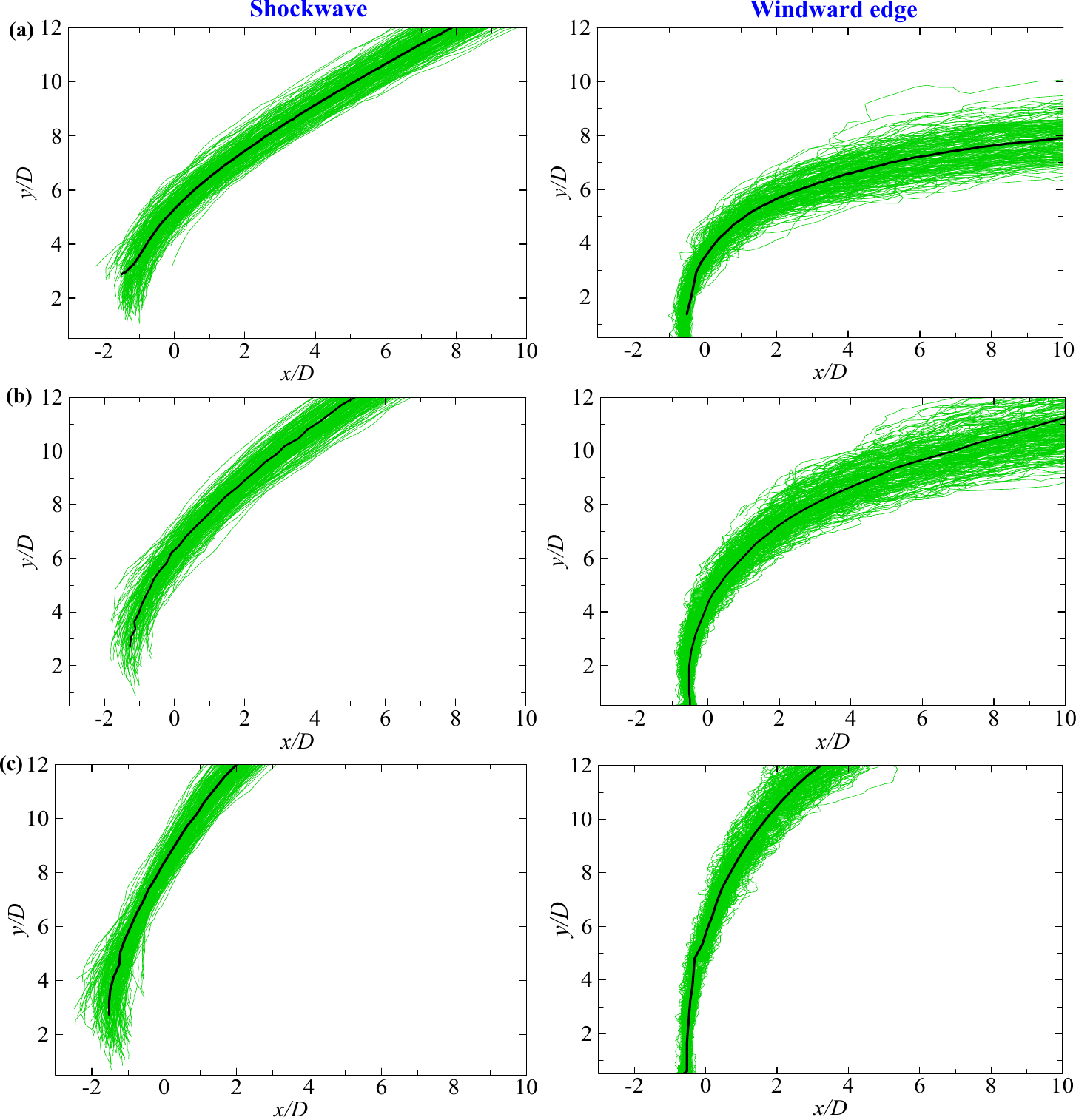}
        \caption{Instantaneous and mean traces of bow shock wave (left) and windward spray edge (right) of liquid jet for (a) $J$ = 1.5, (b) $J$ = 3.7, and (c) $J$ = 9.7, and $AR$ = 3.3. In all the plots, instantaneous traces of the shock wave and the windward spray edge are shown in green. The corresponding mean traces are shown in thick black line curves.}
        \label{fig:16}
\end{figure*}
\begin{figure*}
        \centering
        \includegraphics[width = 0.9\textwidth]{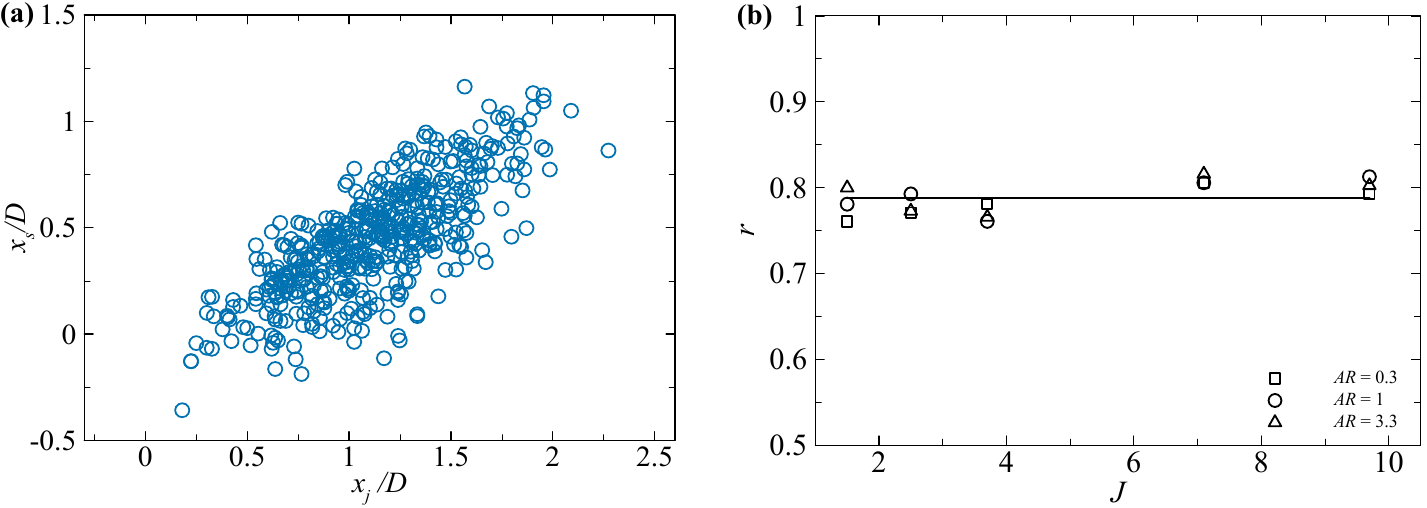}
        \caption{(a) Scatter plot between instantaneous variation in jet and shock positions at $y/D$ = 8 for $J$ = 3.7 and $AR$ = 1. (b) Variation of $r$ with $J$ and for different $AR$. }
        \label{fig:17}
\end{figure*}

The important instantaneous flow features representing this flow-field interaction, viz. incoming boundary layer fluctuations ($u_l$), windward spray edge ($x_j/D$ and $h/D$), and bow shock positions ($x_s/D$ and $y_s/D$), are illustrated in figure~\ref{fig:13}. Using an image detection algorithm for the images (figures \ref{fig:9}, \ref{fig:10}, and \ref{fig:11}), and by capturing the intensity jump across the liquid jet and shock position, the measurements of the instantaneous windward spray edge and the shock position are performed \citep{medipati2023liquid,medipati2025elliptic}. About 1000 images captured during a single experimental run are used for statistical convergence. Figures~\ref{fig:14} (a), (b), and (c) show the large number of instantaneous traces (blue) of bow shockwave (left side) and windward spray edge (right side) for (a) $J$ = 1.5, (b) $J$ = 3.7, and (c) $J$ = 9.7, and for $AR$ = 1, respectively. The thick black curves in all of these plots represent the corresponding mean traces of the shock wave (left) and windward spray edge (right).  As anticipated, as the $J$ increases, the mean liquid jet penetration height increases; correspondingly, the mean shock position is located at a higher upstream position, as seen earlier.  By comparing the quantified blue region relative to its mean shapes (black) of shockwave and windward spray edge for each $J$ case, it is observed that a large-scale unsteadiness is involved during these flow-field interactions. It is evident that the width of the blue region about its mean gradually shrinks as $J$ increases, which is highlighted by a horizontal pink line on all plots at a given cross-streamwise location ($ y/D = 8$). The length of the pink-colored line in the streamwise direction, as shown from top to bottom in the plots, emphasizes a decreasing trend in unsteadiness for both the shock wave (left side) and windward edge (right side).  The similar shrinkage in the width of the orange and green regions about the respective mean as $J$ increases is also observed in the case of  $AR$ = 0.3, and $AR$ = 3.3 as depicted in figure \ref{fig:15} and figure \ref {fig:16}, respectively.  This strongly indicates the decrease in the unsteadiness of the shock wave and windward spray edge with $J$ and is consistent among the orifice $AR$ cases.  Noteworthy to mention that although there is a significant decrease in the fluctuations in liquid jet and shock positions (top portion) with an increase in $J$, the fluctuations in the shock foot remain the same and significant irrespective of $J$ and $AR$.  This is due to the fact that the shock foot is primarily affected by the boundary layer fluctuations (presence of low/high momentum streaks)\citep{munuswamy2022shock,gruber1996bow}.

To further understand the influence of $J$ on the involved unsteadiness in jet penetration height and shock position, along with their interaction dynamics, the instantaneous locations of the shockwave and liquid jet are plotted simultaneously, and a detailed correlation analysis is performed for each $J$ and different $AR$ cases, and is discussed below.

A typical temporal variation of the instantaneous windward spray edge position ($x_j/D$) and the corresponding bow shock wave locations ($x_s/D$) in the streamwise direction is plotted and shown in figure~\ref{fig:17}a. These are shown at a chosen cross-streamwise location of  $y/D$ = 8 for a moderate $J$ case of 3.7 and $AR$ = 1 (see figure \ref{fig:14} b).  It is observed from figure~\ref{fig:17}a that $x_s/D$ (-0.4 $D$ to 1.1 $D$) and $x_j/D$ (0.25 $D$ to 2.25 $D$) vary significantly between the instants, emphasizing that significant unsteadiness is present in both shock and windward edge positions.  At a similar cross-streamwise location ($y/D$ = 8), for a low $J$ case of 1.5 (figure \ref{fig:14} a) and a higher $J$ case of 9.7 (figure \ref{fig:14} c), the shock position oscillates between 1 $D$ to 3 $D$ and -1.2 $D$ to -0.75 D, during which the windward edge fluctuates between 1.8 $D$ to 7.5 $D$ and -1.4 $D$ to 0.3 $D$, respectively.  This implies a significant increase and decrease in unsteadiness, as exhibited by the shock wave and the liquid jet position, due to the decrease and increase of $J$, respectively.  This indicates that $J$ significantly influences the unsteadiness in jet penetration height and shock position, as discussed earlier.  Noteworthy that the range of these values is comparable with the variation in respective mean quantities due to the variation in $J$. The interaction dynamics between the shock and liquid jet positions are further understood by performing a correlation coefficient analysis between $x_s/D$ and $x_j/D$ values for different $J$ and at different $y/D$ locations. The correlation coefficient ($r$) is found to be 0.8 irrespective of $J$, as shown in figure~\ref{fig:17}b.  This implies that, irrespective of $J$, a strong correlation exists between the instantaneous shock and liquid jet positions.  This indicates that at the instant a liquid jet penetrates more deeply into the cross-flow, the shock is displaced further upstream, and vice versa.  These findings align with the qualitative observations discussed earlier.  The chosen transverse location is only representative, and a similar value of $r$ is found at other locations as well.   A similar correlation analysis is also performed for different $J$ by changing the orifice $AR$.  The value of $r$ turns out to be similar to the $AR$ = 1 case, indicating the interaction dynamics between the shock wave and liquid jet positions is independent of $AR$ as well (figure \ref{fig:17}b). The range of values (minimum to maximum) of fluctuations in the shock ($x_s/D$) and liquid jet positions ($x_j/D$) for different $J$ and $AR$ = 1, 0.3, and 3.3, at a sample location ($y/D$ = 8) are shown in the tables \ref{tab:AR1}, \ref{tab:AR0.3}, and \ref{tab:AR3.3}, respectively.  Comparison of these tabulated values across different $J$ and $AR$ shows that there is a significant increase in fluctuations in $x_s/D$ and $x_j/D$ due to the decrease in $J$, and also as the orifice $AR$ changes to 0.3 and 3.3.  This suggests that apart from $J$, the unsteadiness involved in these interactions can be passively altered due to the change in the orifice shape and its orientation with respect to cross-flow.
\begin{table*}
\centering
    \begin{tabular}{c   c   c } 
    \hline
    $J$ & $x_s$ & $x_j$  \\ 
   \hline
     1.5 & 1$D$ to 3$D$ & 1.8$D$ to 7.5$D$   \\
     3.7 & -0.4$D$ to -1.1$D$ & 0.25$D$ to 2.2$D$   \\ 
     9.7 &-1.2$D$ to -0.75$D$  & -1 to 0.3$D$  \\
      \hline
         \end{tabular}
\caption{Temporal variation of shock and jet positions in the streamwise direction at $y/D$ = 8 for different $J$ and $AR$ = 1.}  
     \label{tab:AR1}
\end{table*}
\setlength{\tabcolsep}{10pt} 
\renewcommand{\arraystretch}{1.5} 
\begin{table*}
\centering
    \begin{tabular}{c   c   c } 
    \hline
    $J$ & $x_s$ & $x_j$  \\ 
   \hline
     1.5 & 1.8$D$ to 4.3$D$ & 2.5$D$ to 14$D$   \\
     3.7 & 0.2$D$ to 2.4$D$ & 1$D$ to 5.5$D$   \\ 
     9.7 &0$D$ to -0.9$D$  & 0$D$ to 0.75$D$  \\
      \hline
         \end{tabular}
\caption{Temporal variation of shock and jet positions in the streamwise direction at $y/D$ = 8 for different $J$ and $AR$ = 0.3.}  
     \label{tab:AR0.3}
\end{table*}
\setlength{\tabcolsep}{10pt} 
\renewcommand{\arraystretch}{1.5} 
\begin{table*}
\centering
    \begin{tabular}{c   c   c } 
    \hline
    $J$ & $x_s$ & $x_j$  \\ 
   \hline
     1.5 & 0.8$D$ to 3.8$D$ & 3$D$ to 16$D$   \\
     3.7 & 0.05$D$ to 2.1$D$ & 1$D$ to 5.5$D$   \\ 
     9.7 &-0.8$D$ to 0.4$D$  & 0 to 1.5$D$  \\
      \hline
         \end{tabular}
\caption{Temporal variation of shock and jet positions in the streamwise direction at $y/D$ = 8 for different $J$ and $AR$ = 3.3.}  
     \label{tab:AR3.3}
\end{table*}

\begin{figure*}
        \centering
        \includegraphics[width = 0.9\textwidth]{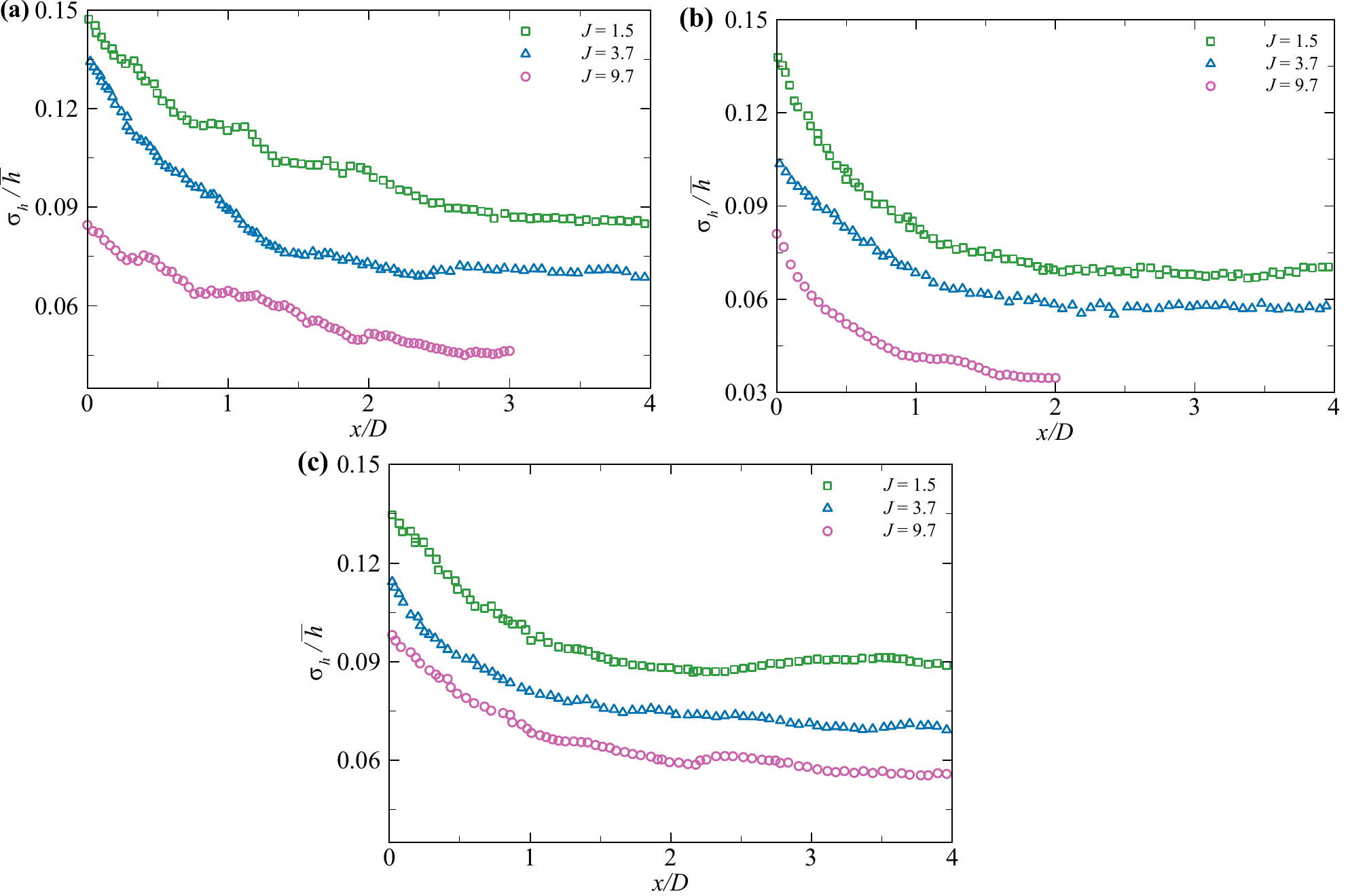}
        \caption{Variation of $\sigma_h$ along the streamwise direction for different $J$ and for (a) $AR$ = 0.3, (b) $AR$ = 1, and (c) $AR$ = 3.3.}
        \label{fig:18}
\end{figure*}

The influence of $J$ on the fluctuations/unsteadiness in the liquid jet penetration height is characterized by quantifying the standard deviation of penetration height ($\sigma_h$) along the streamwise direction.  These values are normalized with the respective local mean penetration height ($\overline{h}$)  and their variation with $x/D$ for different $J$ and for $AR$ = 0.3, 1, and 3.3  are shown in figures \ref{fig:18}a, b, and c, respectively.  For a given $J$, it is observed that along $x/D$ the unsteadiness in penetration height decreases monotonically and reaches a constant value as the liquid jet reaches its maximum penetration height. This trend is consistent with $J$ and $AR$. For a given $AR$, the highest and lowest values of $\sigma_h$/$\overline{h}$ are found to be exhibited by the lower ( = 1.5) and higher ( = 9.7) $J$ cases, respectively.  For example, at $x/D$ = 2, in the case of $AR$ = 0.3 (figure \ref{fig:18}a), the unsteadiness significantly reduces from approximately 11 $\%$ to 4.5 $\%$ when the $J$ has increased from 1.5 to 9.7, and is consistent among the orifice $AR$ cases.  Whereas for a given $J$, the highest and lowest unsteadiness are found to be for $AR$ = 3.3 and 1, respectively. For example, for the $J$ = 9.7 case, at $x/D$ = 4, the unsteadiness considerably reduces from approximately 6 $\%$ to 3 $\%$ when the $AR$ has changed from 3.3 to 1, as can be observed from figures {\ref{fig:18}b and \ref{fig:18}c}.  These disparities in the unsteadiness with $J$ and $AR$ are caused by the fact that the interaction of the liquid jet with the incoming boundary layer streaks is significantly different among different cases, as we shall discuss next.  

\begin{figure*}
        \centering
        \includegraphics[width = 0.9\textwidth]{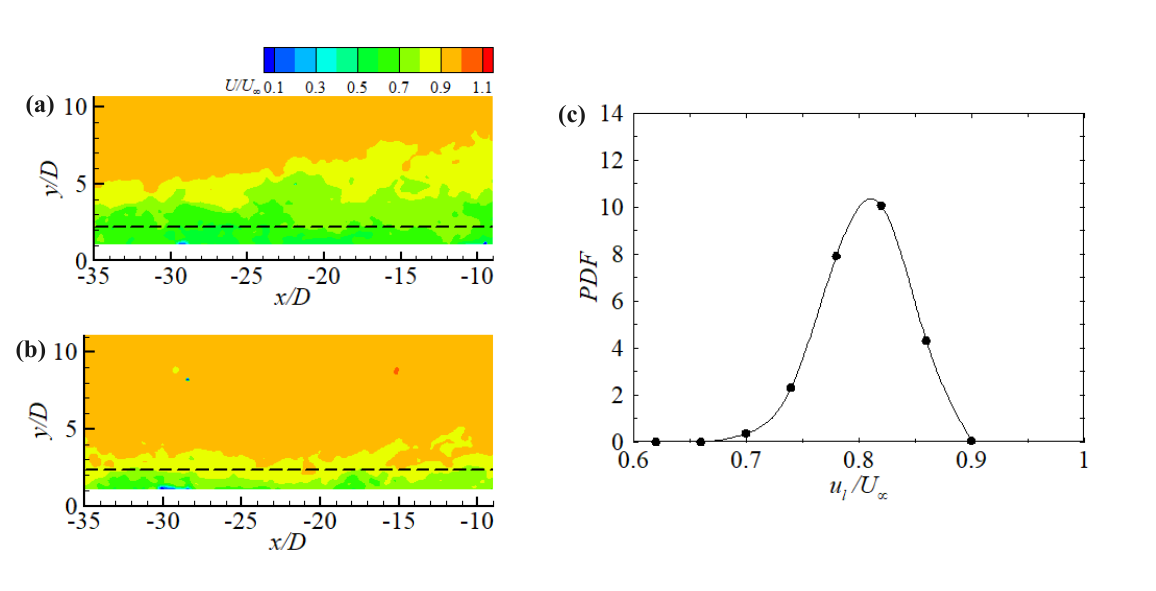}
        \caption{Instantaneous velocity fields of the boundary layer in the upstream of the jet exit in the mid span plane to emphasize the fluctuations in the cross-flow momentum between the instants (a) thicker and (b) thinner boundary layers due to the presence of low and high speed momentum streaks, respectively. The horizontal line indicates the transverse location at which the line-averaged velocity is computed. }
        \label{fig:19}
\end{figure*}

To reveal the source of unsteadiness and its variation with $J$ and $AR$, PIV measurements are conducted in the upstream of the jet. From the visualizations of two extreme cases of $J$  as shown in figures \ref{fig:9}a \& \ref{fig:9}b ($AR$ = 1), \ref{fig:10}a \& \ref{fig:10}b ($AR$ = 0.3), and \ref{fig:11}a \& \ref{fig:11}b ($AR$ = 3.3), it is evident that in the case of $J$ = 1.5,  a large portion of the liquid jet is within the boundary layer and interacting intensely with the incoming boundary streaks. As the $J$ gradually increases to 9.7, the penetration height has significantly increased, and now a large portion of the liquid jet is outside the boundary layer and is interacting predominantly with the outer cross-flow fluid.  This is apparent by comparing the penetration height relative to the incoming boundary layer thickness ($\delta/D$) as marked in the corresponding images.  Hence, a narrow field of view of about 25$D$ by 12$D$ is chosen in the upstream of the jet exit for a detailed PIV.  Two sample instantaneous boundary layer velocity fields captured from PIV are shown in figures~\ref{fig:19}(a) and (b).  These velocity fields between the instants highlight the fact that there is a significant variation in the incoming boundary layer thickness as experienced by the liquid jet; a thicker boundary layer instant is associated with a low-speed momentum streak (figure \ref{fig:19}a), while a thinner boundary layer is associated with a high-speed momentum streak (figure \ref{fig:19}b). To determine this unsteadiness, a line-averaged velocity is computed in the streamwise direction ($u_l/U_\infty$) by considering the streamwise velocity vectors between $x/D$ = -10 and -35, at $y/D$ = 2.2 (or $y/\delta$ = 0.3), which is chosen within the boundary layer as shown schematically in figure \ref{fig:13}.  In the past, the same quantity has been used by many researchers \cite{ganapathisubramani2007effects,murugan2016shock,munuswamy2022shock} to determine the existence of low- and high-speed streaks in the incoming turbulent boundary layer, which can control the instantaneous shock position. It is important to note that the primary difference between the previous \cite{ganapathisubramani2007effects,murugan2016shock,munuswamy2022shock} and the present study is that the shock wave is formed due to the obstruction caused by a solid body in the studies of \cite{ganapathisubramani2007effects,murugan2016shock}, whereas it is due to the liquid and gaseous jets in the current and \cite{munuswamy2022shock} studies, respectively.  Thus, the occurrence of boundary layer streaks can influence the instantaneous jet penetration height in the present study, in addition to the shock foot position. The value of $u_l/U_\infty$ significantly varies between 0.6 and 0.9 as observed from figure \ref{fig:19}c.  This indicates that there is a large variation in the boundary-layer momentum as encountered by the injected liquid jet between the instants.  This reveals that with a thicker boundary layer instant, the existence of a low momentum streak is associated with a lower boundary layer momentum, thus helping the jet to penetrate more into the cross-flow, and vice versa for the thinner boundary layer instant. 

In the case of $J$ = 1.5 (figures \ref{fig:14}a, \ref{fig:15}a, and \ref{fig:16}a), since the jet penetration height is lower and is of the order of mean boundary layer thickness ($\overline{h}/D$  $\sim$ $\delta/D$ ), hence the entire liquid jet is within the boundary layer and is interacting intensely with the oncoming boundary layer streaks,  as discussed earlier.  This has resulted in the formation of highly irregular, larger, and unsteady windward surface waves, leading to the formation of more corrugated shock structures with large variations in time.  This eventually leads to the large-scale unsteadiness in the overall liquid jet penetration height and shock position. In contrast, in the case of $J$ = 9.7 (figures \ref{fig:14}c, \ref{fig:15}c, and \ref{fig:16}c), the injected liquid jet has penetrated significantly into the cross-flow and is significantly higher than the mean boundary layer thickness ($\overline{h}/D$  $>> $$\delta/D$), resulting in a reduced interaction with the oncoming boundary streaks leading to a regular, smaller and steady formation of the windward surface waves, leading to the formation of a relatively smoother shock waves.  This ultimately resulted in lower unsteadiness in the jet penetration height and shock position, and it remains consistent for any orifice aspect ratio ($AR$). 

Typically, the streaks present within the turbulent boundary layer span approximately 40$\delta$ and 0.5$\delta$ in the streamwise and spanwise directions, respectively \citep{ganapathisubramani2007effects}. In the present study, the frontal jet dimension to streak width ratio is significantly different between the orifice $AR$ cases.  These ratios are 0.15, 0.27, and 0.5 for $AR$ = 0.3, 1, and 3.3, respectively.  Among all the $AR$ cases, the frontal to streak width ratio is higher for the $AR$ = 3.3 liquid jets, resulting in an enhanced interaction of the liquid jets with the boundary layer streaks.  Hence, irrespective of $J$, liquid jets injected with the $AR$ = 3.3 orifice have exhibited an enhanced unsteadiness in jet penetration and shock position (figure \ref{fig:18}c). Whereas, in the case of $AR$ = 0.3 (figure \ref{fig:18}a), although the frontal to streak width ratio is the least, the unsteadiness is considerably higher compared to $AR$ = 1 (figure \ref{fig:18}b).  This is most likely due to the increased interaction of the lateral surfaces of the liquid jet with the streaks in the streamwise direction (40$\delta$).          

To summarize, when a liquid jet is injected into a supersonic cross-flow, there is an inherent large-scale unsteadiness associated with the liquid jet penetration height and shock position.  The origin of unsteadiness is due to the large variation in the incoming boundary layer thickness between instants as experienced by the liquid jet.  An instant with a thicker boundary layer, the occurrence of the low-speed momentum streaks within the boundary layer, favors the jet to instantaneously penetrate more into the cross-flow and displaces the bow shock wave to an upstream location.  The opposite scenario occurs when the boundary layer is thinner, as it is associated with a high-speed momentum streak.  It is found that the ratio of the momentum flux of the jet with respect to cross-flow ($J$) plays a significant role in controlling the unsteadiness in jet penetration height and shock position.  In particular, the lower the $J$, the stronger the interaction with the boundary layer streaks, resulting in enhanced unsteadiness in every aspect of the flow-field interactions, while the unsteadiness is reduced at higher $J$ conditions.  These unsteadiness aspects of the flow are further altered due to the difference in the interaction area of the liquid jet with the oncoming boundary streaks caused by the changes in the orifice $AR$.  This near-field unsteadiness in the liquid jet penetration height and shock position, as influenced by $J$ and $AR$, will strongly impart and control the large-scale unsteadiness in the distribution of the droplets across the plume.

\setlength{\tabcolsep}{10pt} 
\renewcommand{\arraystretch}{1.5} 
\section{Unsteady pressure measurements in the liquid injection line}
\label{Unsteady pressure measurements in the liquid injection line}
 The large-scale unsteadiness observed in all aspects of flow features naturally raises a question about the nature of pressure fluctuations in the liquid injection line and whether any dominant frequency is involved during these complex, unsteady interactions.  In this section, we will focus on the impact of incoming boundary layer fluctuations and low-frequency shock oscillations induced by Shock Wave Boundary Layer Interactions (SWBLI) occurring ahead of the liquid jet on pressure fluctuations and the entangled frequencies in the liquid injection line.  The variations in the amplitude of these fluctuations due to the changes in $J$ and $AR$, followed by the details about their spectral content, are discussed.  
 
 To delve into this, an unsteady pressure transducer is introduced in the liquid injection line as discussed in the experimental methods section \ref{Experimental methods}. Detailed measurements are performed both with and without cross-flow conditions for each $J$ and $AR$ case.  Since these experiments are limited by the working pressures (maximum 6.7 bar) of the piezo pressure transducer, these are performed only for the $J$ values of 1.5, 2.5, and 3.3, and for all the $AR$ cases.  We believe these are sufficient to understand the effect of $J$ and $AR$ on the pressure fluctuations.  
 \begin{figure*}
        \centering
        \includegraphics[width = 0.6\textwidth]{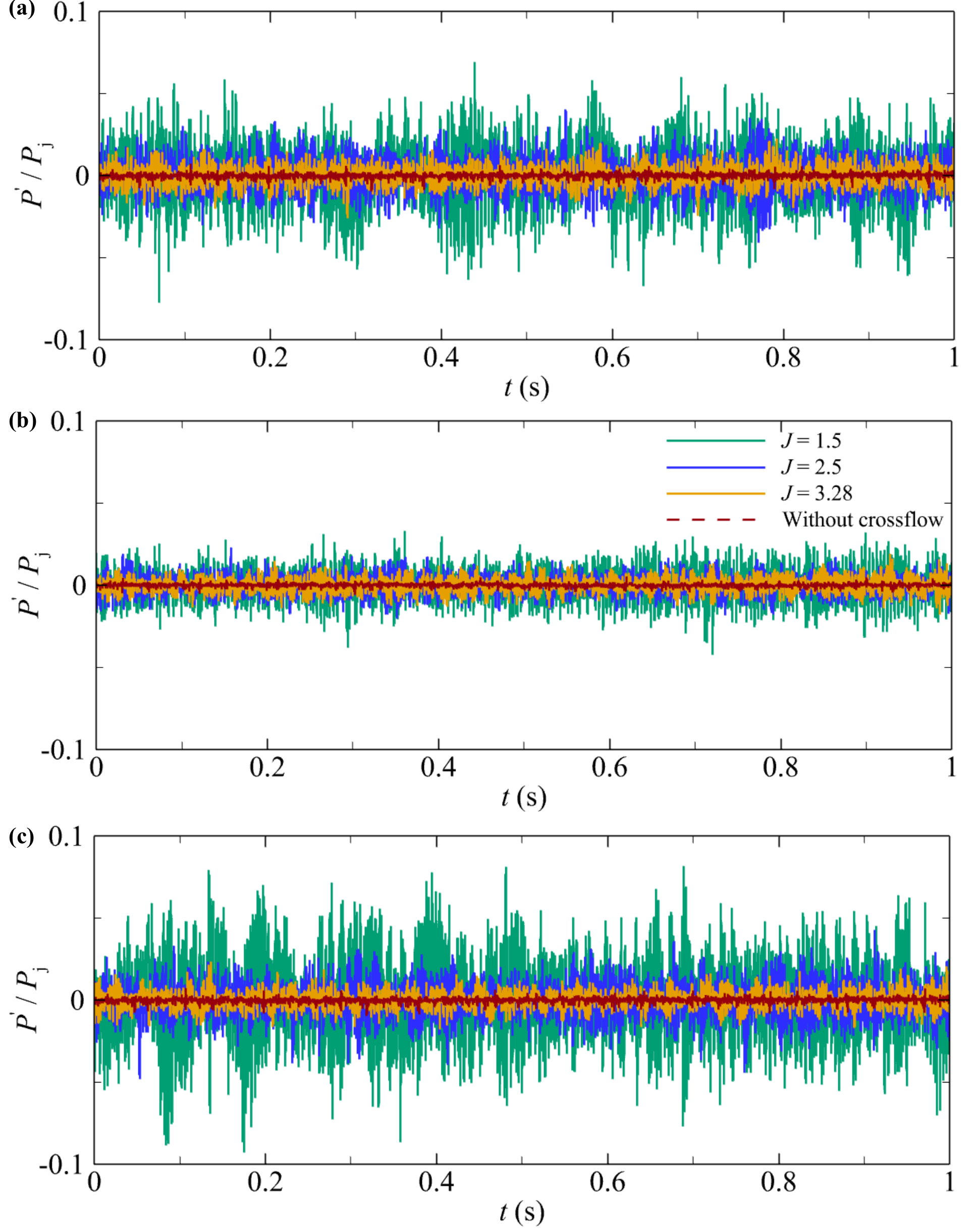}
        \caption{Pressure traces in the liquid injection line corresponding to without and with cross-flow conditions for different $J$ and for (a) $AR$ = 0.3, (b) $AR$ = 1, and (c) $AR$ = 3.3, respectively.  $P_j$ represents the mean injection pressure.}
        \label{fig:20}
\end{figure*}
\begin{figure*}
        \centering
        \includegraphics[width = 0.9\textwidth]{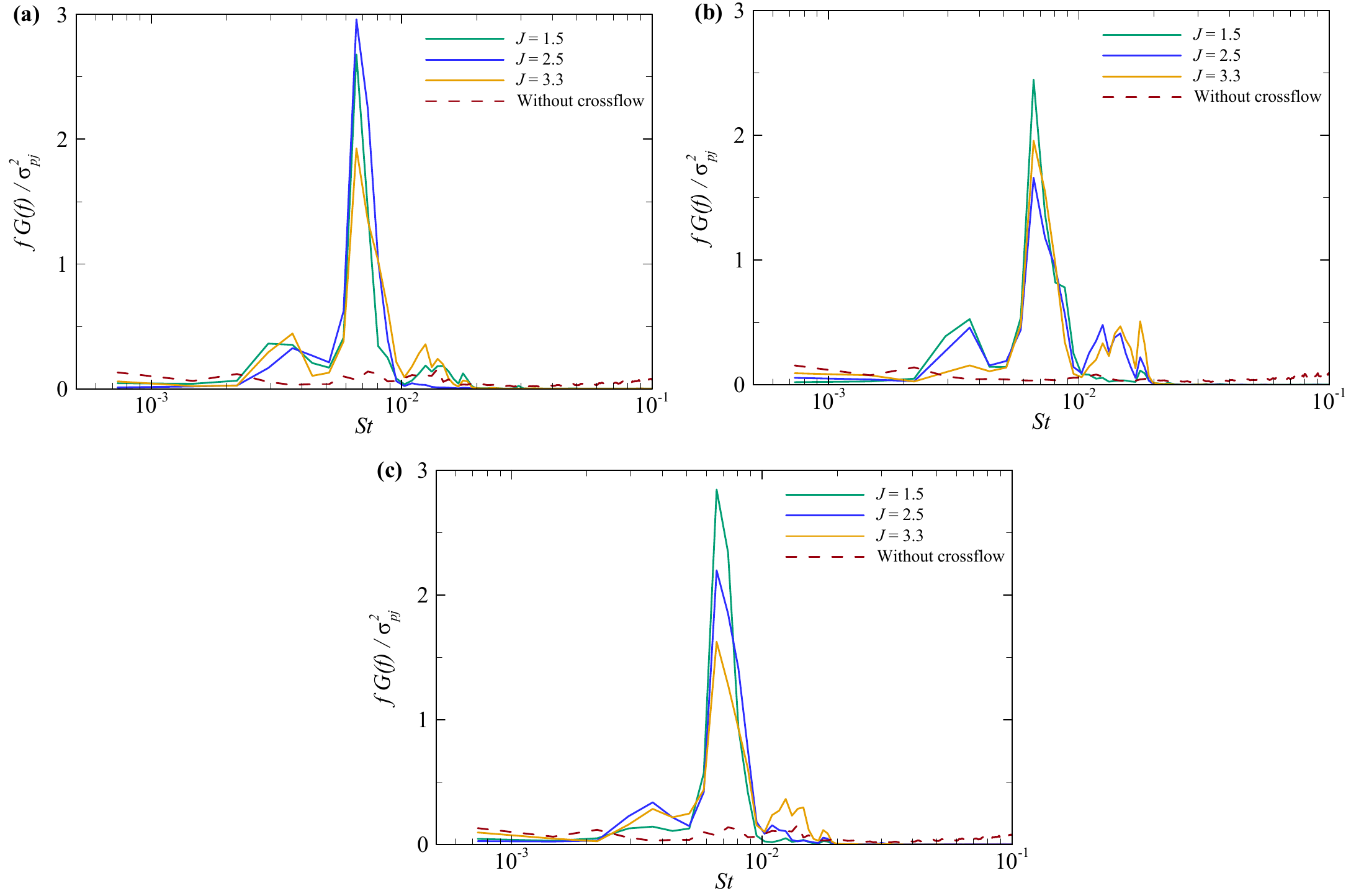}
        \caption{ Premultiplied power spectral density versus Strouhal number corresponding to without and with cross-flow conditions for different $J$ and for (d) $AR$ = 0.3, (e) $AR$ = 1, and (f) $AR$ = 3.3, respectively. $\sigma_{pj}$ represents the rms of the fluctuations of the injection pressure.}
        \label{fig:21}
\end{figure*}

The long-time pressure traces (normalized by the mean injection pressure) of the liquid jet with and without cross-flow conditions for different $J$ are shown in figures \ref{fig:20} (a), (b), and (c), for (a) $AR$ = 0.3, (b) $AR$ = 1, and (c) $AR$ = 3.3, respectively.  The corresponding frequency-multiplied pressure power spectra (normalized by the root-mean-square) are shown in figures \ref{fig:21} (a), (b), and (c) respectively.  First, the experiments are conducted without cross-flow at different injection pressures and with different orifice $AR$s.  The injection pressures are chosen such that the mass flow rate matches the different $J$ cases during the injection with cross-flow.  The liquid jet enters the test-section with no cross-flow present, i.e., the wind tunnel is switched off.  It is evident that the recorded pressure traces (dashed red) are quite steady, with only slight fluctuations observed.  As anticipated, the corresponding power spectra (dashed red) show very low energy levels with the absence of any significant peaks over a range of frequencies ($f$) represented by a dimensionless Strouhal number, $St$ = $f\delta/U_\infty$ ($\delta$ and $U_\infty$ represent the boundary thickness just before the injection location and the free-stream velocity, respectively).  The nature of pressure fluctuations and their spectra in the absence of supersonic cross-flow exhibit similar behavior for a wide range of injection pressures and with different orifice $AR$ cases. 

Once the supersonic flow is switched on, the pressure traces exhibit a dramatically different behavior, as shown in figure~\ref{fig:20} for $J$ = 1.5, 2.5, and 3.3, and for different $AR$ cases. It is evident that the liquid jet exhibits a larger amplitude of the pressure fluctuations, and the amplitude of fluctuations significantly decreases with $J$ for a given orifice $AR$. For example, in the case of $AR$ = 1 (figure \ref{fig:20} (b)), the amplitude of fluctuations has decreased from 3$\%$ to 1 $\%$ while $J$ has increased from 1.5 to 3.3.  The same behavior is also seen in the case of $AR$ = 0.3 (figure \ref{fig:20} (a)) and 3.3 (figure \ref{fig:20} (c)) cases. This confirms that the unsteadiness in the liquid jet has significantly reduced as the $J$ increases due to the reduced interaction with the oncoming boundary layer streaks, which is in close agreement with the analysis pertaining to measurements of the unsteadiness in the liquid jet penetration height and shock position, as discussed in section~\ref{Unsteady behavior of penetration height and shock position}.  Apart from the variation with $J$, the pressure fluctuations also showed a dependency on orifice $AR$. It is found that for a given $J$, the $AR$ = 3.3 has exhibited a larger amplitude of fluctuations, followed by $AR$ = 0.3, and then by $AR$ = 1. This trend also indicate that the liquid jets injected through an orifice with a higher frontal to streak width ratio ($AR$ = 3.3) interacts more with the incoming boundary layer streaks, therefore exhibiting larger unsteadiness, and is also in compliance with the effect of $AR$ on the unsteadiness in the liquid jet penetration height and shock position, as discussed in section \ref{Unsteady behavior of penetration height and shock position} and in our previous study \citep{medipati2025elliptic}. A closer look at these pressure traces has shown that a definite frequency is involved in all the cases of the $J$ and $AR$.  Hence, the corresponding long-time spectra exhibit a clear dominant peak in the energy centered at $St$ $\approx$ 0.007, and is invariant with $J$ and $AR$ (figures \ref{fig:21}a, b, and c). It is important to note that the chosen definition of $St$ is widely used in the SWBLI literature. The observed peak in energy at $St$ $\approx$ 0.01 is consistent with the low-frequency shock oscillations as discussed in detail in the review article of \cite{clemens2014low}.  This confirms that the source of the frequency of oscillations of the liquid jet is due to the strong SWBLI between the shock formed ahead of the liquid jet and the incoming boundary layer, and the origin of unsteadiness is the variation in the incoming boundary layer thickness/momentum between the instants, similar to the \cite{ganapathisubramani2007effects,murugan2016shock,munuswamy2022shock}.  In addition, the pressure fluctuations in the liquid injection line indicate fluctuations in the injected mass flow, which are self-induced due to the naturally occurring SWBLI upstream of the jet. 

Further, the invariance in the spectra for different $J$ suggests the asymptotic nature of the flow field upstream of the jet.  To confirm this, we take data from studies of the flow field upstream of a cylinder in supersonic flow \citep{westkaemper1968turbulent,sedney1977separation}. In these studies, an important upstream quantity, the mean separation distance ($S/D$), which carries the dynamics of the flow field, is measured for varying  cylinder height ($h/D$) and the free-stream Mach number. These studies revealed that for $M_\infty$ = 2.5, the value of $S/D$ remained at a constant value ($\approx$ 2.2) for $h/D$ > 3 and 0.38 $\le$ $h/\delta$ $\ge$ 4.05 (see fig~2 of \citep{sedney1977separation}). In the current study, the liquid jet mean penetration height ($\overline{h}/D$) as controlled by $J$ is closely inline with the cylinder height in the above studies \cite{westkaemper1968turbulent,sedney1977separation}.  Hence, one would expect a similar value of constant $S/D$ in the present study ($AR$ = 1) for all the $J$ cases since the mean penetration height, $\overline{h}$, is found to be 8$D$ for the present lower $J$ case (= 1.5), which is significantly higher compared to height of the cylinder ($h$ = 3$D$) \citep{sedney1977separation}. This comparison suggests the existence of an asymptotic flow-field behavior upstream of the jet for the range of $J$ values chosen in the current study, and hence no significant variation in the peak of the spectral content with $J$.  Also, there is no variation seen in the location of peak with a change in orifice $AR$, which is also aligned with the general collapse of the spectra in the various interaction types of existing SWBLI studies (see fig.3d of \citep{sedney1977separation} and \citep{gonsalez1993correlation}).  This indicates a general similarity in the interaction dynamics and physics of SWBLI in the upstream region of the liquid jet over a wide range of $J$ and $AR$.       

\section{Conclusions}
\label{Conclusions}
In the present study, we have experimentally investigated elliptical liquid jets in a supersonic cross-flow with a free-stream Mach number, $M_\infty$ of 2.5.  The main focus of the study is to understand the influence of the momentum flux ratio ($J$) on the formation of surface waves at the liquid-gas interface, the breakup mechanism, shock structures, unsteadiness in the liquid jet penetration height and shock position, the pressure fluctuations in the liquid injection line, and the effect of SWBLI on the liquid jet frequencies.  These experiments were conducted for a wide range of $J$ values varying between 1.5 and 9.7. Three orifice aspect ratios ($AR$ = $b/a$), namely 0.3, 1, and 3.3, were also varied for each $J$, where $AR$ is defined as the ratio of orifice spanwise to streamwise dimensions. 

Liquid jets, when injected into a supersonic cross-flow, are at high risk of instability due to the intense acceleration of the liquid jet by the supersonic air. This results in the formation of the Rayleigh-Taylor instability (RTI) on the windward liquid-gas interface \cite{taylor1950instability,medipati2025elliptic}. In the present study, it is found that $J$ has a significant influence on the formation of windward surface waves. In the lower $J$ (= 1.5) case, the windward surface waves formed are larger in wavelength ($\lambda/b$) due to the lower acceleration of the liquid jet. This is due to the higher jet deflection into the cross-flow, resulting in lower drag forces experienced by the liquid jet. Furthermore, the formation of surface waves is highly irregular and unsteady in nature. This is due to the stronger interaction with the incoming boundary layer streaks.  The immediate consequence of this is the formation of a highly corrugated bow shock wave upstream of the jet, with large variations between instants. This leads to large variations in the velocity field encountered by the liquid jet. This kind of behavior is further intensified in the case of $AR$ = 0.3, due to the fact that the frontal area of the liquid jet is lower (resulting in lower drag), which leads to a further reduction in the acceleration of the liquid jet, and consequently, the formation of larger wavelength RT waves. This, in turn, results in the formation of more corrugated shock structures with much stronger variations in time. This acts as a source of strong unsteadiness in the downstream spray core \citep{medipati2025elliptic}. 

In contrast, as the $J$ increases gradually (penetration height increases), the formation of surface waves is relatively more regular and steady. The value of $\lambda/b$ gradually decreases with $J$, with the least value occurring for the largest $J$ = 9.7 case. This is due to the increase in the acceleration of the liquid jet resulting from the higher drag forces experienced by the liquid jet, which is caused by the lower jet deflection into the cross-flow. This leads to the formation of a stronger bow shock wave with significantly reduced corrugations, along with reduced temporal variations. This results in relatively lower variation in the velocity field ahead of the liquid jet. Furthermore, this behavior is more pronounced in the case of $AR$ = 3.3, due to the higher frontal area of the liquid jet, which results in higher accelerations of the liquid jet (higher drag). Thus, the RT waves are very small, and the formation of the shockwave is free from corrugations. This ultimately results in reduced unsteadiness in the downstream spray core \citep{medipati2025elliptic}. It may be noted that the variation in experimental values of $\lambda/b$ with $J$ and $AR$ captures a similar decreasing trend with the theoretical predictions of the RT wavelength derived from the linear instability analysis pertaining to the maximum growth rate \citep{chandrasekhar2013hydrodynamic}. 

However, irrespective of $J$, in the case of $AR$ = 0.3 and 1, the dominant atomization mechanism of the liquid jet is through the formation of Kelvin-Helmholtz instability (KHI) on the lateral surfaces due to the intense shear between the liquid jet and high speed air \citep{chai2015numerical,xiao2016large}, resulting in vigorous mass stripping leading to the formation of liquid ligaments/tongues, which further undergo atomization due to RTI. This is closely in line with the well-studied coaxial liquid jet atomization  \citep{varga2003initial}. This is more intensified in the case of the $AR$ = 0.3 due to the presence of large lateral surfaces. Furthermore, as $J$ increases, the primary atomization mechanism through KHI $\rightarrow$ RTI will begin to occur at higher transverse heights. This is most likely due to the formation of stronger shock waves, resulting in a lower free-stream velocity after the shock, and hence, a lower shear experienced by the liquid jet. This can result in a significant difference in the amount of mass stripping from the jet surface and can eventually influence the dispersion of droplets across the spray plume. In contrast, in the case of $AR$ = 3.3 (irrespective of $J$), RT waves are smaller and the atomization mechanism is purely through RTI, resulting in the catastrophic breakup of the liquid jet.

Further, the breakup time scales are computed for each $J$ and $AR$ case. It is found that an increase in $J$ results in a lower breakup time due to the enhanced accelerations of the liquid jet (increase in drag), indicating early breakup of the liquid jet. In addition, as the $AR$ increases, it results in a further enhancement of the acceleration of the liquid jet. Hence, the jet breakup times are further lower.     

It is found that there is large-scale unsteadiness in the liquid jet penetration height and shock position in lower $J$ cases, irrespective of the $AR$. As $J$ increases, the unsteadiness in all aspects of the flow field decreases correspondingly.  The same behavior was also observed in the pressure fluctuations in the liquid injection line.  PIV measurements upstream of the liquid jet revealed a significant variation in the incoming boundary layer thickness, indicating the existence of low-speed/high-speed momentum streaks whose streamwise and spanwise length scales are 40$\delta$ and 0.5$\delta$, respectively \citep{ganapathisubramani2007effects}. This results in a significant variation in the boundary layer momentum (total cross-flow momentum) between the instances, as encountered by the injected liquid jet.  In the case of lower $J$, a large portion of the liquid jet is within the boundary layer and intensely interacts with the boundary layer streaks, resulting in larger unsteadiness. Further, it is found that the unsteadiness is dependent on orifice $AR$. Since in the case of $AR$ = 3.3, the frontal dimension to streak width ratio is highest, the unsteadiness is higher, followed by $AR$ = 0.3 and then by $AR$ = 1, which is consistent with \cite{medipati2025elliptic}. 

Power spectral densities of the liquid jet during the interaction with the supersonic cross-flow exhibited higher energy levels with a dominant peak in the pressure signal centered at $St$ = $f\delta/U_\infty$ $\approx$ 0.007, which is in close agreement with the low-frequency shock oscillations of SWBLI studies \citep{clemens2014low}. The dominant peak in $St$ is independent of $J$ and $AR$, implying a general similarity of the flow-field upstream of the jet, similar to protuberances in supersonic flow \citep{westkaemper1968turbulent,sedney1977separation}. 
 
To conclude, the present experimental investigation reveals that the momentum flux ratio ($J$) significantly influences various aspects of the flow-field interaction, in addition to the orifice $AR$ effects studied previously \citep{medipati2025elliptic}. This implies that both $J$ and $AR$ play a crucial role in determining flow-field interactions, which can have significant consequences in applications where such configurations are employed.

\section*{Acknowledgements}
The authors acknowledge support from the Joint Advanced Technology Programme (JATP) at IISc (Grant reference: SCRAMSPRAY (JATP 0158)) for financial support of the experimental work.

\section*{Data Availability Statement}
The data from this study are available upon request.

\nocite{*}
\bibliography{aipsamp}

\end{document}